\documentclass[aps,pre,onecolumn,superscriptaddress,longbibliography]{revtex4-2}
\usepackage{amsmath}
\usepackage{amsfonts}
\usepackage{amssymb}
\usepackage{lmodern,dsfont}
\usepackage{graphicx}
\usepackage[usenames,dvipsnames]{xcolor}
\usepackage{bm}
\usepackage[english=american]{csquotes}
\usepackage{xr}

\newcommand{\Av}[1]{\left\langle #1 \right\rangle}
\newcommand{\av}[1]{\langle #1 \rangle}
\newcommand{\n}{\nonumber}

\newcommand{\artanh}{\text{artanh}}
\renewcommand{\eqref}[1]{Eq.~(\ref{#1})}

\begin{document}
\author{Andreas Dechant}
\affiliation{Department of Physics \#1, Graduate School of Science, Kyoto University, Kyoto 606-8502, Japan}

\title{Minimum entropy production, detailed balance and Wasserstein distance for continuous-time Markov processes}
\date{\today}

\begin{abstract}
We investigate the problem of minimizing the entropy production for a physical process that can be described in terms of a Markov jump dynamics.
We show that, without any further constraints, a given time-evolution may be realized at arbitrarily small entropy production, yet at the expense of diverging activity.
For a fixed activity, we find that the dynamics that minimizes the entropy production is given in terms of conservative forces.
The value of the minimum entropy production is expressed in terms of the graph-distance based Wasserstein distance between the initial and final configuration.
This yields a new kind of speed limit relating dissipation, the average number of transitions and the Wasserstein distance.
It also allows us to formulate the optimal transport problem on a graph in term of a continuous-time interpolating dynamics, in complete analogy to the continuous space setting.
We demonstrate our findings for simple state networks, a time-dependent pump and for spin flips in the Ising model.
\end{abstract}

\maketitle

\section{Introduction and overview}

Entropy production quantifies how much a physical system is driven out of equilibrium. 
In practice, entropy production manifests itself in the form of dissipation, that is, energy that is irreversibly lost into the environment. 
Thus, minimizing the entropy production for a physical process is highly desirable from a practical point of view.
But minimum entropy production is also interesting from a more fundamental point of view.
There is a long-standing history of attempts to characterize non-equilibrium steady states via a minimum entropy production principle, see Ref.~\cite{Jay80} for a historical and Ref.~\cite{Mae13} for a more recent overview.

In this work, we consider the minimization of entropy production on a theoretical level, but motivated from the practical point of view.
The general question we want to address is, given a physical process, what is the minimum amount of entropy production that is required to realize it?
Obviously, in order to answer this question, we need to specify some details about the process and what are the parameters of the minimization.
We will focus on Markovian dynamics in continuous time, for which entropy production can be defined, at least mathematically, in an unambiguous manner.
For this class of dynamics, we can further differentiate between diffusion processes, where the state space is continuous, and jump processes, where the state space is discrete.
In most cases, the continuous state space of diffusion processes is the $d$-dimensional real space $\mathbb{R}^d$ and the dynamics can mathematically be described in terms of a Langevin equation \cite{Cof17} for the degrees of freedom or a Fokker-Planck equation \cite{Ris86} for their probability density.
A physical process may then be characterized by the time-evolution of the probability density, or, if only the initial and final configuration are of interest, the initial and final probability density.
In the latter case, it has been shown in Refs.~\cite{Aur11,Aur12,Dec19b}, that the problem of minimizing the entropy production during the process is equivalent to calculating the Wasserstein distance between the two probability densities.
The Wasserstein distance is a concept from optimal transport theory \cite{Vil08}, which characterizes the optimal way of transforming an initial configuration into a final one.
Since the Wasserstein distance characterizes the minimum entropy production, it also provides a lower bound on the entropy production \cite{Dec19b}s. 
More recently, this approach has also been applied to the case where the detailed process instead of just the initial and final configurations is specified \cite{Nak21,Dec21}.

For the case of a jump process, a similar formulation has only been developed very recently \cite{Van21}.
There, it was shown that a quantity resembling the Wasserstein distance, which is obtained by solving a modified optimal transport problem, provides a lower bound on the entropy production.
However, this approach has two issues: First, the definition of the Wasserstein-distance-like quantity depends on the process itself and thus, unlike the Wasserstein distance in the continuous case, is not just a function of the initial and final configuration; it is further not directly related to the types of Wasserstein distance that are typically considered in the mathematical literature.
Second, the argument of Ref.~\cite{Van21} only applies to systems whose transition rates satisfy detailed balance and thus does not allow for non-conservative forces.
The latter point is important in light of another recent work \cite{Rem21}, where it was shown that the entropy production is generally minimized by rates that do not satisfy detailed balance.
This is in contrast to the continuous case, where the force that realizes minimum entropy production is always conservative and thus satisfies detailed balance \cite{Mae14,Dec21}.

In this work, we address both of these issues.
In Ref.~\cite{Rem21} the entropy production was minimized under the condition that the symmetric part of the transition rates is fixed.
We show that by adopting a slightly weaker condition, minimum entropy production does indeed correspond to rates that satisfy detailed balance.
Specifically, instead of the symmetric part of the rates, we fix the activity, which can be interpreted as the overall rate at which transitions occur in the system.
Crucially, the minimum entropy production can be expressed in terms of the usual, graph-distance based Wasserstein distance, which only depends on the connectivity of the state space and the initial and final configuration.
This generalizes the results of Refs.~\cite{Aur11,Aur12,Dec19b,Nak21,Dec21} to the case of a discrete state space.

Our paper is structured as follows:
We consider a Markovian dynamics on a discrete state space, that is characterized by the time-dependent occupation probabilities $\bm{p}_t = (p_t(1),\ldots,p_t(N))$ of the states labeled by $1,\ldots,N$.
This dynamics is introduced in detail in Section \ref{sec-markov-jump}.
For a given time evolution of the occupation probabilities $\bm{p}_t$, we want to know is the minimum entropy production compatible with this time-evolution.
As we discuss in section \ref{sec-mod-rates}, without any further constraints, the answer to this problem is trivial: we can make the entropy production arbitrarily small, so there is no finite lower bound.
However, the price we have to pay for this is that the overall rate of transitions between states, quantified by the activity, diverges as the entropy production tends to zero.
By contrast, as we show in Section \ref{sec-wasserstein}, there exists a minimum value of the activity, that is, a given time-evolution requires a certain overall transition rate.
The main result of this section is that the minimum activity can be expressed in terms of the usual Wasserstein distance on a graph between the initial and final occupation probability.
While the minimum activity requires a diverging entropy production, we show in Section \ref{sec-minent} that this formalism can also be used to construct the dynamics that minimize the entropy production for a given activity.
This allows us to express the minimum entropy production in terms of the Wasserstein distance.
These concepts are generalized to finite time-differences in Section \ref{sec-finite-time}, where, instead of the entire time-evolution, we only specify the initial and final occupation probabilities.
This leads to a new type of speed limit for the transition between two states.
In Section \ref{sec-benamou}, we discuss the relation between a discrete and continuous state state space and derive a generalization of the Benamou-Brenier formalism to the former case.
Section \ref{sec-examples} discusses the application of the results to a few simple state networks and to the Ising model.

\section{Continuous-time Markov processes and detailed balance} \label{sec-markov-jump}
We consider a continuous-time Markovian dynamics on a set of $N$ states.
Transitions between states $j$ and $i$ occur at a rate $W_t(i,j) \geq 0$ which may depend on time $t$ through some external protocol.
Then, the probability $p_t(i)$ of being in state $i$ evolves according to the master equation \cite{Van92,Gar02}
\begin{align}
d_t p_t(i) = \sum_j \big( W_t(i,j) p_t(j) - W_t(j,i) p_t(i) \big) \label{master} ,
\end{align}
with a prescribed set of initial occupation probabilities $\bm{p}_0 = (p_0(1),\ldots,p_0(N))$.
In the following, we assume that the rates satisfy the condition $W_t(i,j) = 0 \Leftrightarrow W_t(j,i) = 0$, i.~e.~that there are no unidirectional transitions.
Then, we can parameterize the rates as \cite{Rem21}
\begin{align}
W_t(i,j) = k(i,j) \omega_t(i,j) \exp\bigg(\frac{1}{2} A_t(i,j) \bigg), \label{rates} 
\end{align}
with $k(i,j) = k(j,i) \in \lbrace 0,1 \rbrace$, $\omega_t(i,j) = \omega_t(j,i) \geq 0$ and $A_t(i,j) = -A_t(j,i)$.
Physically, $k(i,j)$ encodes the connectivity of the state network, i.~e., between which states transitions are allowed ($k(i,j) = 1$) or forbidden ($k(i,j) = 0$).
The parameters $\omega_t(i,j)$ constitute the symmetric part of the transition rates, which we can interpret as energetic barriers between the states. 
By contrast, the antisymmetric part $A_t(i,j)$ can be interpreted as the forces driving transitions between the states.
Note that the parameterization \eqref{rates} can represent an arbitrary set of transition rates, specifically, the symmetric and antisymmetric part are determined from the rates as
\begin{align}
\omega_t(i,j) &= \sqrt{W_t(i,j) W_t(j,i)}, \qquad 
 A_t(i,j) = \ln \bigg(\frac{W_t(i,j)}{W_t(j,i)} \bigg) .
\end{align}
Note that, in the following sections, we will sometimes omit the subscript $t$ in favor of a more compact notation when we consider the dynamics only at a specific time $t$.
In general, the dynamics described by \eqref{master} are irreversible: Let us consider a trajectory $\Gamma = \lbrace i(t) \rbrace_{t \in [0,\tau]}$, i.~e.~we trace the instantaneous state of the system over a time-interval of length $\tau$.
Then, we can define a probability $P(\Gamma)$ of observing a specific trajectory.
We can also imagine the time-reversed version of the dynamics, in which the system starts from the final state of \eqref{master}, $\bm{p}_0^\dagger = \bm{p}_\tau$, and evolves according to the time-reversed protocol $W^\dagger_t(i,j) = W_{\tau - t}(i,j)$.
The probability of observing the time-reversed trajectory $\Gamma^\dagger = \lbrace i(\tau-t) \rbrace_{t \in [0,\tau]}$ in the time-reversed dynamics is given by $P^\dagger(\Gamma^\dagger)$.
The irreversibility of the system is measured by the entropy production \cite{Sei12}
\begin{align}
\Delta S^\text{irr} = \sum_{\Gamma} P(\Gamma) \ln \bigg(\frac{P(\Gamma)}{P^\dagger(\Gamma^\dagger)} \bigg) \label{entropy-kl} .
\end{align}
Mathematically, this is the Kullback-Leibler divergence between the path probabilities of the forward and reverse dynamics; this quantity is positive and vanishes only if, for every trajectory, the probability of observing its time-reversed version in the time-reversed dynamics is the same as for the forward trajectory in the original dynamics.
In terms of the occupation probabilities and rates, the entropy production can be expressed as
\begin{align}
\Delta S^\text{irr} &= \int_0^\tau dt \ \sigma_t \qquad \text{with} \label{entropy} \qquad 
\sigma_t = \frac{1}{2} \sum_{i,j} \big(W_t(i,j) p_t(j) - W_t(j,i) p_t(i) \big) \ln \bigg(\frac{W_t(i,j) p_t(j)}{W_t(j,i) p_t(i)} \bigg) ,
\end{align}
where $\sigma_t \geq 0$ is the rate of entropy production.
The entropy production rate only vanishes, if for every pair of states $(i,j)$, the probability of a transition from $j$ to $i$ is precisely the same for as the reverse transition,
\begin{align}
W_t(i,j) p_t(j) = W_t(j,i) p_t(i) \label{db},
\end{align}
which is the detailed balance condition and equivalent to the system being in equilibrium \cite{Van92}.
From \eqref{master}, this immediately implies $d_t \bm{p}_t = 0$ and thus, a vanishing entropy production rate can only occur if the occupation probabilities are independent of time, which is called a steady state.
However, a steady state does not imply detailed balance; a steady state with a constant, non-vanishing rate of entropy production is referred to as a non-equilibrium steady state.
We assume that, fixing $t = \hat{t}$, the dynamics described by \eqref{master} reaches a steady-state $\bm{p}_{\hat{t}}^\text{st}$ in the long-time limit.
The detailed balance condition \eqref{db} in the steady state then is equivalent to
\begin{align}
\exp \big(A_{\hat{t}}(i,j) \big) = \exp \big( \phi^\text{st}_{\hat{t}}(j) - \phi^\text{st}_{\hat{t}}(i) \big) \label{db-rates} ,
\end{align}
where we defined $\phi^\text{st}_{\hat{t}}(i) = - \ln (p_{\hat{t}}^\text{st}(i))$.
This means that the forces $A_t(i,j)$ have to derive from a potential function $\phi_t(i)$ and are thus conservative.
If $A_t(i,j)$ cannot be written in this way, that is, for non-conservative forces, detailed balance is broken and the steady state is out of equilibrium.
We remark that both \eqref{db} and \eqref{db-rates} are called detailed balance condition, the difference being that for \eqref{db} the system is already in the equilibrium state, whereas \eqref{db-rates} only implies that an equilibrium state exists and will eventually be reached if the rates are kept constant.
For convenience, we refer to \eqref{db-rates} as the detailed balance condition in the following.

\section{Time evolution at vanishing entropy production} \label{sec-mod-rates}
From \eqref{entropy} it is obvious that the entropy production rate depends on both the transition rates and the occupation probabilities.
These dependencies can be made explicit by writing \cite{Sei12}
\begin{align}
\sigma_t &= \sigma_t^\text{m} + \sigma_t^\text{s} \qquad \text{with} \label{entropy-medium-shannon} \\
\sigma_t^\text{m} &= \sum_{i,j} W_t(i,j) p_t(j) \ln \bigg(\frac{W_t(i,j)}{W_t(j,i)} \bigg) \qquad \text{and} \qquad \sigma_t^\text{s} = - \sum_{i} \ln \big(p_t(i) \big) d_t p_t(i) . \n
\end{align}
The first term is generally referred to as the rate of entropy production in the medium; it quantifies the asymmetry in the transition rates.
The second part is the rate of change of the Gibbs-Shannon entropy of the system; it depends explicitly on the time-evolution of the probabilities.
The central question that we wish to answer in this work is, given a time-evolution of the occupation probabilities $\lbrace \bm{p}_t \rbrace_{t \in [0,\tau]}$, what is the minimum value of the entropy production compatible with this time-evolution.
The motivation for this question is twofold:
On the one hand, we may think of the evolution $\bm{p}_t$ as the desired effect of driving the system, for example, when we want to pump the system into a higher-energy state.
Form this point of view, we want to minimize the dissipation associated with the driving.
On the other hand, since any change in the probabilities $\bm{p}_t$ implies a non-vanishing entropy production, we may ask how much we can infer about the dissipation in the system, which is often challenging to measure directly, from the observed evolution of $\bm{p}_t$.
As it turns out, without any further qualifiers, the answer to the above question is trivial: The minimum entropy production associated with the time-evolution $\bm{p}_t$ is zero.
To see this, we define a new set of transition rates $W^\mathcal{A}_t(i,j)$, which leads to the same time-evolution at an arbitrary entropy production.
First, we define currents $J_t(i,j)$ and traffic $D_t(i,j)$ as
\begin{align}
J_t(i,j) &= W_t(i,j) p_t(j) - W_t(j,i) p_t(i) \label{current},  \qquad
D_t(i,j) = W_t(i,j) p_t(j) + W_t(j,i) p_t(i) .
\end{align}
$J_t(i,j)$ can be interpreted as the directed current across the transition from $j$ to $i$, while $D_t(i,j)$ measures the rate of transitions along the transition in either direction.
We then define the rates
\begin{align}
W_t^\mathcal{A}(i,j) = \frac{1}{2 p_t(j)} \Big( \mathcal{A} + \text{sign}\big(J_t(i,j)\big) \Big) \big\vert J_t(i,j) \big\vert \label{rates-mod} ,
\end{align}
with some real parameter $\mathcal{A} \geq 1$.
A straightforward computation shows that the resulting currents are given by
\begin{align}
J_t^\mathcal{A}(i,j) = J_t(i,j) .
\end{align}
So, the rates \eqref{rates-mod} leave the currents invariant.
Since, via \eqref{master}, we have $d_t p_t(i) = \sum_j J_t(i,j)$, this also means that the time-derivative of the probabilities, and thus their time-evolution, remains unchanged.
However, the entropy production rate associated with \eqref{rates-mod} is
\begin{align}
\sigma_t^\mathcal{A} = \frac{1}{2}\sum_{i,j} \big\vert J_t(i,j) \big\vert \ln \bigg(\frac{\mathcal{A}+1}{\mathcal{A}-1} \bigg) .
\end{align}
It is easily seen that, for given currents, this can take all values between zero (for $\mathcal{A} \rightarrow \infty$) to positive infinity (for $\mathcal{A} \rightarrow 1$).
For large $\mathcal{A}$, in particular, we have
\begin{align}
\sigma_t^\mathcal{A} \simeq \mathcal{A}^{-1} \sum_{i,j} \big\vert J_t(i,j) \big\vert + O(\mathcal{A}^{-3}) .
\end{align}
Thus, we can make the entropy production rate arbitrarily small without altering the time-evolution of the occupation probabilities.
However, there is a price we have to pay for this:
If we consider the activity, which is defined as
\begin{align}
\chi_t = \sum_{i,j \neq i} W_t(i,j) p_t(j) \label{activity}
\end{align}
and measures the overall rate of transitions in the system, we find
\begin{align}
\chi_t^\mathcal{A} = \frac{\mathcal{A}}{2} \sum_{i,j} \big\vert J_t(i,j) \big\vert.
\end{align}
Thus, even as the entropy production vanishes, the overall rate of transitions in the system diverges, with the product of the two quantities approaching a constant value
\begin{align}
\lim_{\mathcal{A} \rightarrow \infty} \sigma_t^\mathcal{A} \chi_t^\mathcal{A} = \frac{1}{2} \bigg(\sum_{i,j} \big\vert J_t(i,j) \big\vert \bigg)^2 .
\end{align}
We remark that at the special value of the parameter $\mathcal{A}$ at which the activity resulting from \eqref{rates-mod} is the same as the activity in the original system,
\begin{align}
\mathcal{A}^* = \frac{[D_t]}{[|J_t|]} \qquad \text{with} \qquad [D_t] \equiv \sum_{i,j \neq i} D_t(i,j) ,
\end{align}
the corresponding entropy production is a guaranteed lower bound on the entropy production in the original dynamics,
\begin{align}
\sigma_t^{\mathcal{A}^*} = \frac{1}{2} [|J_t|] \ln \bigg( \frac{[D_t]+[|J_t|]}{[D_t]-[|J_t|]} \bigg) \leq \sigma_t \label{entropy-bound-2} .
\end{align}
This inequality can be proven using elementary inequalities, see Appendix \ref{app-bound}.
This means that, using \eqref{rates-mod}, we have a dynamics with the same currents, time-evolution and activity as the original dynamics, yet at a reduced entropy production rate.
The fact that we can reduce the entropy production at a given activity poses the question of whether \eqref{rates-mod} already constitutes the optimum or whether the entropy production may be reduced further.

\section{Wasserstein distance and minimal activity}  \label{sec-wasserstein}
Before we address the minimization of the entropy production, we are going to introduce the concept of Wasserstein distance on a graph.
The intuitive idea of optimal transport \cite{Vil03,Vil08} is to find the optimal way of transforming a probability vector $\bm{q}$ into $\bm{p}$, minimizing the cost of transporting probability in the process.
If the cost of transporting probability is given by the distance over which the probability is transported, then the minimal cost is called Wasserstein distance.
In the present setting, the connectivity matrix $k(i,j)$, which specifies which transitions are allowed or forbidden, is the adjacency matrix of a graph, whose vertices $v_i$ are the states $i \in 1,\ldots,N$ \cite{Wes96}.
If $k(i,j) = 1$, then the vertices $v_i$ and $v_j$ are connected by an edge $e(i,j)$, if $k(i,j) = 0$, then there is no edge between $v_i$ and $v_j$.
For now, we consider a simple undirected graph, that is $e(i,j)$ and $e(j,i)$ are equivalent and there is at most one edge between any pair of states.
We call the collection of vertices and edges the graph $\mathcal{G}(v,e)$.
We further define the graph distance $d_\text{G}(i,j)$ as the minimum number of edges in any path from $v_i$ to $v_j$.
Note that the shortest path between two vertices is generally not unique, so that there may be multiple shortest paths with length $d_\text{G}(i,j)$.
This function is satisfies all the axioms of a distance, $d_\text{G}(i,j) \geq 0$ with equality if and only if $v_i = v_j$, $d_\text{G}(i,j) = d_\text{G}(j,i)$ and $d_\text{G}(i,k) \leq d_\text{G}(i,j) + d_\text{G}(j,k)$ \cite{Wes96}.
We assume that the graph is connected, that is
In terms of this distance, the Wasserstein distance of order 1 or Kantorovich-Rubinstein distance \cite{Vil08} between two probability vectors $\bm{p}$ and $\bm{q}$ is defined as
\begin{align}
\mathcal{W}(\bm{p},\bm{q}) = \text{inf}_{\Pi} \sum_{i,j} d_\text{G}(i,j) \Pi(i,j) \label{wasserstein} .
\end{align}
The infimum is taken over all couplings $\Pi$ between $\bm{p}$ and $\bm{q}$, which satisfy
\begin{align}
\Pi(i,j) \geq 0, \quad \sum_{i} \Pi(i,j) = q(j), \quad \sum_{j} \Pi(i,j) = p(i)  \label{marginals}.
\end{align}
This means that we are looking for the joint probability with marginals $\bm{p}$ and $\bm{q}$ that minimizes the expectation of the graph distance.
As the name implies, the Wasserstein distance is a proper distance, satisfying the corresponding axioms \cite{Vil08}.
An important property of \eqref{wasserstein} is that, in the above setting, the infimum always exists, however, the optimal coupling $\Pi^*$ that realizes the minimal value is not unique.
One reason is that the shortest path between any two vertices is not necessarily unique, so that we have several equivalent ways of transporting probability between the two vertices.
Another reason is that, even if the shortest path is unique, we can easily construct a coupling that leads to the same value.
To do so, assume we have two vertices with $\Pi(k,l) > 0$, which implies that we are moving some probability from $l$ to $k$.
Then, we consider one of the (possibly many) shortest paths from $l$ to $k$ which we denote by $\pi(k,l)$.
By definition $\pi(k,l)$ consists of $d_\text{G}(k,l) \equiv m$ edges.
We denote the directed edge from $j$ to $i$ by $\vec{e}(i,j)$ and write $\pi(k,l) = (\vec{e}(k,j_{m-1}), \vec{e}(j_{m-1},j_{m-2}), \ldots, \vec{e}(j_1,l))$.
Now, we define a new coupling by setting $\widetilde{\Pi}(k,l) = 0$ and $\widetilde{\Pi}(i,j) = \Pi(i,j) + \Pi(k,l)$ for all $\vec{e}(i,j) \in \pi(k,l)$.
Intuitively, this means that, instead of moving probability directly from $v_l$ to $v_k$, we first move it from $v_l$ to $v_{j_1}$, then from $v_{j_1}$ to $v_{j_2}$ and so on, finally arriving at $v_{k}$.
Since, along the shortest path, the graph distance is additive, we are moving the same amount of probability by the same distance and thus the value of \eqref{wasserstein} is the same for $\Pi$ and $\widetilde{\Pi}$.
This means that for any coupling $\Pi$ (which does not have to be the optimal one), we can always construct an equivalent coupling that is non-zero only the edges of the graph.
In the following, we can thus assume that $\Pi(i,j) = 0$ whenever the edge $\vec{e}(i,j)$ is not part of the graph.

Now we make the connection to the Markov jump dynamics \eqref{master}.
We specialize \eqref{wasserstein} to the case $\bm{p} = \bm{p}_{t+dt}$ and $\bm{q} = \bm{p}_t$, that is, we consider the solution of \eqref{master} at two infinitesimally different times.
From \eqref{marginals}, we then obtain the condition
\begin{align}
d_t p(i) = \frac{p_{t+dt}(i) - p_t(i)}{dt} = \frac{1}{dt} \sum_j \big( \Pi(i,j) - \Pi(j,i) \big) \label{master-coupling} .
\end{align}
Since, by the above discussion, we can assume that $\Pi(i,j) > 0$ only if $k(i,j) = 1$ ($W_t(i,j) > 0$) we may thus identify
\begin{align}
\Pi(i,j) = W_t(i,j) p_t(j) dt \label{rates-coupling} .
\end{align}
This provides a one-to-one correspondence between the off-diagonal elements of the coupling and the transition rates.
\eqref{marginals} further fixes the diagonal elements as
\begin{align}
\Pi(i,i) = \bigg(1 - \sum_{k \neq i} W_t(k,i) dt \bigg) p_t(i) ,
\end{align}
which is positive for sufficiently small $dt$.
We can thus compute the activity associated with a given coupling,
\begin{align}
\chi_t &= \sum_{i,j \neq i} W_t(i,j) p_t(j) \label{act-coupling} = \frac{1}{dt} \sum_{i,j \neq i} \Pi(i,j) = \frac{1}{dt} \sum_{i,j} d_\text{G}(i,j) \Pi(i,j) .
\end{align}
In the last step, we used that all pairs of vertices for which $\Pi(i,j) > 0$ we have either $d_\text{G}(i,j) = 1$ (if $i \neq j$) or $d_\text{G}(i,j) = 0$ (if $i = j$).
Apart from the factor $dt$, this is precisely the same as the functional in \eqref{wasserstein}.
Thus, minimizing the activity for a given connectivity and time-evolution is precisely the same as computing the Wasserstein distance between $\bm{p}_t$ and $\bm{p}_{t+dt}$ and
\begin{align}
\chi_t^* = \frac{1}{dt} \mathcal{W}(\bm{p}_{t+dt},\bm{p}_t) \label{minact} .
\end{align}

Next, we want to exploit the known properties of the optimal coupling $\Pi^*$.
A central result of optimal transport theory is the Kantorovich duality \cite{Vil08}.
For the present case, this result states that there exists a function $\psi(i)$ defined on the vertices that satisfies
\begin{align}
\psi(j) - \psi(i) = d_\text{G}(i,j) \quad \text{for all} \; (i,j) \; \text{with} \; \Pi^*(i,j) > 0 .
\end{align}
In other words, for all edges on which the optimal coupling is non-zero, the graph distance can be written in terms of a potential function $\psi(i)$.
Since we may assume that the optimal coupling is non-zero only on the edges of the graph, this implies
\begin{align}
\psi(j) - \psi(i) = 1  \quad \text{for all} \; (i,j) \; \text{with} \; \Pi^*(i,j) > 0 \label{coupling-landscape} ,
\end{align}
that is, we may characterize the optimal coupling by a potential landscape with energy difference $\pm 1$ between neighboring sites.
More precisely, we have $\psi(j) - \psi(i) = 1$ if probability is transported from $j$ to $i$ ($\Pi^*(i,j) > 0$) and $\psi(j) - \psi(i) = -1$ if probability is transported from $i$ to $j$ ($\Pi^*(j,i) > 0$), so that flow of probability described by the optimal coupling is always downhill in the potential.
This also implies that for given $i$ and $j$, probability flows between them in only one direction and the transition rates \eqref{rates-coupling} defined by $\Pi^*$ are unidirectional $W_t^*(i,j) > 0 \ \Rightarrow W_t^*(j,i) = 0$.
From the point of view of minimizing the activity, this is reasonable, as any reverse transitions also contribute to the activity.
However, this also means that at the minimal activity, the entropy production rate diverges.
Since our original goal was minimizing the entropy production, it seems we have only achieved the opposite.
However, as we will see in the next section, the information about dynamics realizing minimal activity will actually be useful for minimizing the entropy production.
We remark that, if the optimal coupling $\Pi^*$ is known, the energy landscape $\psi(i)$ is determined uniquely up to a constant shift \cite{Vil08}.
Conversely, the knowing the energy landscape also determines the Wasserstein distance, which can be written as
\begin{align}
\mathcal{W}(\bm{p}_\tau,\bm{p}_0) = \sum_i \psi(i) \big( p_{t+dt}(i) - p_t(i) \big) .
\end{align}


\section{Minimum entropy production at fixed activity}  \label{sec-minent}
Minimizing the entropy production rate requires minimizing \eqref{entropy} with respect to the transition rates under the constraints that (i) the rates lead to the correct time-evolution \eqref{master} and that (ii) the resulting activity \eqref{activity} has the same value as in the original dynamics.
To simplify the notation we introduce the parameters $C(i,j) = -C(j,i)$ via
\begin{align}
A(i,j) = 2 C(i,j) + \ln \bigg(\frac{p(i)}{p(j)} \bigg) \label{a-c-relation} ,
\end{align}
and omit the subscript $t$ in the following.
In terms of these parameters, the master equation, entropy production rate and activity can be written as
\begin{subequations}
\begin{align}
d_t p(i) &= 2 \sum_j \sqrt{p(i) p(j)} k(i,j) \omega(i,j) \sinh(C(i,j)), \label{c-master} \\
\sigma &= 2 \sum_{i,j} \sqrt{p(i) p(j)} k(i,j) \omega(i,j) C(i,j) \sinh(C(i,j)), \label{c-entropy} \\
\chi &= \sum_{i,j \neq i} \sqrt{p(i) p(j)} k(i,j) \omega(i,j) \cosh(C(i,j)) \label{c-activity} .
\end{align} \label{c-equations}%
\end{subequations}
Then, we minimize \eqref{c-entropy} with respect to $\omega(i,j)$ and $C(i,j)$ under the constraints \eqref{c-master} and \eqref{c-activity}.
Taking the derivative of \eqref{c-equations} with respect to $\omega(i,j)$ and $C(i,j)$ yields the conditions for a stationary point,
\begin{subequations}
\begin{align}
k(i,j) \big( C(i,j) \sinh(C(i,j)) + \gamma \cosh(C(i,j)) \big) &= k(i,j) \big(\lambda(j) - \lambda(i) \big) \sinh(C(i,j)), \label{c-condition-1} \\
k(i,j) \omega(i,j)  \big( C(i,j) \cosh(C(i,j)) + (\gamma+1) \sinh(C(i,j)) \big) &= k(i,j) \omega(i,j) \big(\lambda(j) - \lambda(i) \big) \cosh(C(i,j)) \label{c-condition-2} ,
\end{align}
\end{subequations}
where $\gamma$ and $\lambda(i)$ are Lagrange multipliers.
For all $(i,j)$ such that $k(i,j) \omega(i,j) > 0$, this results in the condition
\begin{align}
\big(\tanh(C(i,j))\big)^2 = \frac{\gamma}{1+\gamma} \equiv q^2 \quad \text{with} \quad 0 \leq q \leq 1 .
\end{align}
Thus for all pairs of states with a non-zero transition rate between them, $C(i,j)$ has to be constant, specifically
\begin{align}
C(i,j) = s(i,j) \artanh(q)
\end{align}
with $s(i,j) = -s(j,i) \in \lbrace -1, 1 \rbrace$.
Plugging this into \eqref{c-condition-1} yields
\begin{align}
\lambda(j) - \lambda(i) &= s(i,j) \mathcal{E}_q \qquad \text{with} \qquad \mathcal{E}_q  = \bigg( \artanh(q) + \frac{q}{1-q^2} \bigg) .
\end{align}
Thus the Lagrange multipliers $\lambda(i)$ define an energy landscape, in which any two states with a non-vanishing probability flow between them are separated by an energy $\mathcal{E}_q$.
This is precisely the condition that is satisfied by the optimal transport coupling $\Pi^*$!
We choose
\begin{align}
\lambda(i) &= \mathcal{E}_q \psi(i), \label{minent-rates} \qquad  \omega(i,j) = \frac{\sqrt{1-q^2}}{2 q dt} \frac{\Pi^*(i,j) + \Pi^*(j,i)}{\sqrt{p(i) p(j)}} .
\end{align}
For this choice, we obtain from \eqref{c-equations}
\begin{subequations}
\begin{align}
d_t p(i) &= \frac{1}{dt} \sum_j \big( \Pi^*(i,j) - \Pi^*(j,i) \big), \\
\sigma &= \frac{2 \artanh(q)}{dt} \sum_{i,j \neq i} \Pi^*(i,j), \\
\chi &= \frac{1}{q dt} \sum_{i,j \neq i} \Pi^*(i,j) .
\end{align}
\end{subequations}
The first equation is satisfied because of \eqref{master-coupling} and the third equation can be solved for $q$,
\begin{align}
q = \frac{\chi^*}{\chi} ,
\end{align}
where we used \eqref{act-coupling} and \eqref{minact}.
This yields the minimal entropy production rate
\begin{align}
\sigma^* = 2 \chi^* \artanh \bigg(\frac{\chi^*}{\chi} \bigg) \label{minent} .
\end{align}
In summary, for a given connectivity of the state network $k(i,j)$ and time-evolution $\bm{p}_t$, there exists a minimal activity $\chi^*$, which is equal to the Wasserstein distance between $\bm{p}_t$ and $\bm{p}_{t+dt}$.
The minimal entropy production rate at a given activity $\chi$ is then determined by the ratio of $\chi^*$ and $\chi$ and diverges logarithmically as $\chi \rightarrow \chi^*$.
We remark that the above is different from the approach pursued in Ref.~\cite{Rem21}, where the symmetric part $\omega(i,j)$ of the rates was kept fixed.
Intuitively, fixing the activity is a weaker constraint compared to fixing the symmetric parts of all the rates.
Consequently, as we discuss in Appendix \ref{app-minent}, we expect that the minimum entropy production \eqref{minent} is smaller than the minimum entropy production of Ref.~\cite{Rem21} in most cases, even though this relation is not strict.

The somewhat convoluted notation of the above derivation does not lend itself to physical intuition.
However, we can use \eqref{rates-mod} to gain a more intuitive understanding.
Suppose that, in \eqref{rates-mod}, instead of the original rates $W(i,j)$, we use the rates $W^*(i,j)$ corresponding to the optimal coupling $\Pi^*$ via \eqref{rates-coupling}.
Since the rates $W^*(i,j)$ are unidirectional, we then find
\begin{align}
W^\mathcal{A}(i,j) = \left\lbrace \begin{array}{ll} \frac{\mathcal{A}+1}{2} W^*(i,j) &\text{if} \; W^*(i,j) > 0 \\[2 ex]
\frac{\mathcal{A}-1}{2} W^*(j,i) &\text{if} \; W^*(j,i) > 0 . \end{array} \right. \label{rates-mod-optimal}
\end{align}
The corresponding entropy production rate and activity are given by
\begin{subequations}
\begin{align}
\sigma^\mathcal{A} &= 2 \chi^* \artanh \bigg(\frac{1}{\mathcal{A}} \bigg) \\
\chi^\mathcal{A} &= \mathcal{A} \chi^* .
\end{align}\label{minent-mod}%
\end{subequations}
For $\mathcal{A} = 1$, we re-obtain the unidirectional rates corresponding to minimal activity and diverging entropy production.
For any $\mathcal{A} > 1$, however, the reverse transition rate becomes non-zero and thus the entropy production rate is finite.
Thus \eqref{rates-mod-optimal} corresponds to a symmetrized version of the unidirectional rates $W^*(i,j)$.
For the choice $\mathcal{A} = \chi/\chi^*$, the activity is equal to the activity of the original dynamics and the entropy production yields the minimal value at this activity.
In that sense, the minimal entropy production is obtained by first solving the optimal transport problem and then symmetrizing the resulting transition rates by just the right amount.
In terms of \eqref{rates}, the symmetric and antisymmetric part of the transition rates realizing the minimal entropy production rate are given by
\begin{subequations}
\begin{align}
\omega^*(i,j) &= \sqrt{\bigg(\frac{\chi}{\chi^*} \bigg)^2 - 1} \frac{\Pi^*(i,j) + \Pi^*(j,i)}{2 \sqrt{p(i) p(j)} dt} \\
A^*(i,j) &=  \phi(j) - \phi(i) + 2 \artanh\bigg(\frac{\chi^*}{\chi}\bigg) \big(\psi(j) - \psi(i) \big) ,
\end{align} \label{rates-mod-optimal-split}%
\end{subequations}
with $\phi(i) = -\ln p(i)$.
This form explicitly shows that the forces $A^*(i,j)$ are derived from a potential and thus satisfy the detailed balance condition \eqref{db}.
This potential includes two contributions:
The term involving $\phi(i)$ derives from differences between the instantaneous occupation probabilities; it favors transitions from states with high probability to states with low probability.
The second term involving $\psi(i)$ derives from the structure of the optimal coupling \eqref{coupling-landscape}.
Since $\psi(j) - \psi(i) = \pm 1$, we may interpret the prefactor as the energy scale defining the energy landscape.
As we approach the minimal activity, $\chi \rightarrow \chi^*$, the energy scale diverges, while the symmetric part $\omega(i,j)$ tends to zero, which reproduces the finite unidirectional transition rates corresponding to the solution of the optimal transport problem.

We remark that we can also calculate the pseudo-entropy production rate \cite{Shi21}, which is defined by
\begin{align}
\rho = \sum_{i,j} \frac{\big(W(i,j) p(j) - W(j,i) p(i) \big)^2}{W(i,j) p(j) + W(j,i) p(i)} \label{pseudo-def}.
\end{align}
Like the entropy production, the pseudo-entropy production quantifies the breaking of detailed balance in the system.
While it cannot be expressed in terms of the time-forward and time-reversed path probabilities like \eqref{entropy-kl}, it has the advantage that it remains finite in the presence of unidirectional transitions.
It is further a lower bound on both the entropy production rate and the activity: $\rho_t \leq \sigma_t$ and $\rho_t \leq 2 \chi_t$, where the former bound turns into an equality close to equilibrium.
For \eqref{rates-mod-optimal}, the pseudo-entropy production rate is given by
\begin{align}
\rho^\mathcal{A} = \frac{2 \chi^*}{\mathcal{A}} .
\end{align}
Solving \eqref{c-equations} by minimizing $\rho$ instead of $\sigma$ yields precisely the same conditions, so that the entropy production rate and pseudo-entropy production rate are simultaneously minimized for $\mathcal{A} = \chi/\chi^*$,
\begin{align}
\rho^* = \frac{2 (\chi^*)^2}{\chi} \label{pseudo-min} .
\end{align}

\section{Finite times}  \label{sec-finite-time}
In the derivation of the minimum entropy production rate \eqref{minent}, we focused on the situation where the entire evolution of the probabilities is specified.
However, in many situations, only the initial and final configuration $\bm{p}_0$ and $\bm{p}_\tau$ are of interest.
In this case, instead of the instantaneous activity $\chi_t$, it is more natural to fix its time-integral
\begin{align}
\int_0^\tau dt \ \chi_t = \Av{M} \label{transition-number},
\end{align}
which is equal to the average number of transitions in the time interval $[0,\tau]$.
Then, we want to minimize the total entropy production \eqref{entropy} during the time $\tau$, keeping the average number of transitions fixed.
As before, we first establish some properties of the optimal transport problem.
Suppose we know an optimal coupling $\Gamma^*$ between $\bm{p}_0$ and $\bm{p}_\tau$.
Then, from \eqref{marginals}, we have
\begin{align}
p_\tau(i) - p_0(i) = \sum_{j} \big( \Gamma^*(i,j) - \Gamma^*(j,i) \big) .
\end{align}
Then, we define
\begin{align}
\Pi^*(i,j) = \left\lbrace \begin{array}{ll}
\Gamma^*(i,j) \frac{dt}{\tau} &\text{for} \; i \neq j \\[2ex]
1 - \sum_{k} \Gamma^*(k,i) \frac{dt}{\tau} &\text{for} \; i = j .
\end{array} \right.
\end{align}
This satisfies
\begin{align}
\frac{p_\tau(i) - p_0(i)}{\tau} = \frac{1}{dt} \big(\Pi^*(i,j) - \Pi^*(j,i) \big) \label{master-coupling-constant}. 
\end{align}
and, for sufficiently small $dt/\tau$, $\Gamma^*(i,j) > 0$.
If we can find a dynamics that satisfies
\begin{align}
\bm{p}_t &= \bm{p}_0 + \frac{t}{\tau} \big(\bm{p}_\tau - \bm{p}_0\big) \quad \Rightarrow \quad d_t \bm{p}_t = \frac{\bm{p}_\tau - \bm{p}_0}{\tau}  \label{time-derivative-constant},
\end{align}
that is, whose solution interpolates between $\bm{p}_0$ and $\bm{p}_\tau$ at a constant rate of change of the probability, then, by \eqref{master-coupling-constant}, $\Pi^*$ is a coupling between $\bm{p}_t$ and $\bm{p}_{t+dt}$.
We further have
\begin{align}
\mathcal{W}(\bm{p}_\tau,\bm{p}_0) = \sum_{i,j} d_\text{G}(i,j) \Gamma^*(i,j) = \frac{\tau}{dt} \sum_{i,j} d_\text{G}(i,j) \Pi^*(i,j) .
\end{align}
Suppose we divide the time interval $[0,\tau]$ into $K \gg 1$ intervals $[k dt,(k+1)dt]$.
We can write the above as
\begin{align}
\mathcal{W}(\bm{p}_\tau,\bm{p}_0) = K \sum_{i,j} d_\text{G}(i,j) \Pi^*(i,j) \label{wasserstein-split} .
\end{align}
Because the Wasserstein distance satisfies the triangle inequality, we also have
\begin{align}
\mathcal{W}(\bm{p}_\tau,\bm{p}_0) \leq K \mathcal{W}(\bm{p}_{t+dt},\bm{p}_t) \label{wasserstein-triangle} ,
\end{align}
where we used that the distance between $\bm{p}_t$ and $\bm{p}_{t+dt}$ only depends on the (constant) rate of change $d_t \bm{p}_t$ because of \eqref{master-coupling}.
Comparing \eqref{wasserstein-split} and \eqref{wasserstein-triangle}, we then have
\begin{align}
\sum_{i,j} d_\text{G}(i,j) \Pi^*(i,j) \leq \mathcal{W}(\bm{p}_{t+dt},\bm{p}_t) .
\end{align}
Since $\Pi^*$ is a coupling between $\bm{p}_t$ and $\bm{p}_{t+dt}$, it is already is an optimal one, since, by definition, the Wasserstein distance is equal to the infimum of the quantity on the left.
This means that for a constant rate of change of the probability, \eqref{time-derivative-constant}, we have
\begin{align}
\frac{\mathcal{W}(\bm{p}_\tau,\bm{p}_0)}{\tau} = \frac{\mathcal{W}(\bm{p}_{t+dt},\bm{p}_t)}{dt} .
\end{align}
In other words, a constant rate of probability change between the initial and final state yields the geodesic between the two states with respect to the metric defined by the Wasserstein distance.
The coupling $\Pi^*$ also defines the rates that lead to the dynamics \eqref{time-derivative-constant} via \eqref{rates-coupling}.

The above discussion shows that the optimal process that minimizes the total number of transitions is one whose rate of change in the probability is constant.
For this process, we have
\begin{align}
\av{M}^* = \tau \chi^*_t = \tau \frac{\mathcal{W}(\bm{p}_{t+dt},\bm{p}_t)}{dt} = \mathcal{W}(\bm{p}_\tau,\bm{p}_0).
\end{align}
That is, the minimal number of transitions required to transform the initial state $\bm{p}_0$ into the final state $\bm{p}_\tau$ is given by the Wasserstein distance between the two states.
This provides a direct interpretation of the Wasserstein distance between two states in terms of the process connecting the two states.
Further, since $\chi^*_t$ and $\chi_t$ are independent of time, we obtain from \eqref{minent}
\begin{align}
\Delta S^{\text{irr},*} = 2 \mathcal{W}(\bm{p}_\tau,\bm{p}_0) \artanh \bigg(\frac{\mathcal{W}(\bm{p}_\tau,\bm{p}_0)}{\av{M}} \bigg) \label{minent-finite} .
\end{align}
Thus, the minimal amount of entropy production that is required to transform $\bm{p}_0$ into $\bm{p}_\tau$ is given in terms of the Wasserstein distance and the average number of transitions.
The proof that a constant rate of change in the probability indeed minimizes the entropy production is given in Appendix \ref{app-finite}.
For an arbitrary dynamics, the right-hand side \eqref{minent-finite} constitutes a lower bound on the entropy production during the process.
Since the number of transitions is in often proportional to time $\av{M} = \tau \bar{\chi}$, with the time-averaged activity $\bar{\chi}$, \eqref{minent-finite} also translates into a speed limit for the transition from $\bm{p}_0$ to $\bm{p}_\tau$,
\begin{align}
\tau \geq \frac{\mathcal{W}(\bm{p}_\tau,\bm{p}_0)}{\bar{\chi} \tanh \Big( \frac{\Delta S^{\text{irr}}}{2 \mathcal{W}(\bm{p}_\tau,\bm{p}_0)} \Big)} \label{speed-limit} .
\end{align}
That is, the minimal time for a transition between two states is determined by the Wasserstein distance, the activity and the amount of dissipation.
Using the inequalities $\mathcal{W}(\bm{p}_\tau,\bm{p}_0) \geq \delta(\bm{p}_\tau,\bm{p}_0)$ with the total variation distance $\delta(\bm{p},\bm{q}) = \sum_i \vert p(i) - q(i) \vert/2$ and $\tanh(x) \leq x$ we re-obtain the speed limit derived in Ref.~\cite{Shi18},
\begin{align}
\tau \geq \frac{2 \delta(\bm{p}_\tau,\bm{p}_0)^2}{\bar{\chi} \Delta S^\text{irr}} \label{speed-limit-2} .
\end{align}
However, \eqref{speed-limit} is tighter than this bound.
In particular, since the hyperbolic tangent approaches unity for large arguments, we see that in the limit of large dissipation, the activity becomes the limiting factor for the transition time,
\begin{align}
\tau \geq \frac{\mathcal{W}(\bm{p}_\tau,\bm{p}_0)}{\bar{\chi}} .
\end{align}
This means that, in contrast to what is suggested by \eqref{speed-limit-2}, strong driving and large dissipation do not allow us to realize arbitrarily fast transition times.
Another appealing feature of \eqref{speed-limit} is that its derivation shows that the speed limit is tight, that is, there always exists a process that realizes the minimal transition time.
Instead of a speed limit, \eqref{minent-finite} may also be interpreted as a tradeoff relation between dissipation and precision.
Since the Wasserstein distance is equal to the minimum number of transitions, we may write \eqref{minent-finite} as
\begin{align}
\Delta S^{\text{irr}} \geq 2 \av{M}^* \artanh \bigg(\frac{\av{M}^*}{\av{M}} \bigg) \label{precision-tradeoff}.
\end{align}
If the dynamics transforms the initial into the final state with the minimum number of transitions, it has perfect precision: Transitions only occur in the \enquote{correct} direction and never in reverse.
\eqref{precision-tradeoff} implies that this requires infinite dissipation.
If we sacrifice precision and allow transitions to also occur in the reverse direction, then the dissipation becomes finite, however, we also require more transitions to realize the same change in the state.
We remark that similar tradeoff relations have been obtained for Brownian clocks \cite{Bar16} and stochastic currents \cite{Pie18}.
The relation \eqref{precision-tradeoff} establishes a tradeoff relation on the basis of the system state itself, rather than derived quantities.
In addition to the lower bound corresponding to \eqref{speed-limit-2}, we can also give an upper bound on the minimum entropy production.
The maximal graph distance between any two vertices is defined as the diameter of the graph $\max_{(i,j) \in \mathcal{G}} d_\text{G}(i,j) = d(\mathcal{G})$ \cite{Wes96}, which also is an upper bound on the Wasserstein distance \cite{Vil08}.
We thus obtain lower and upper bounds on the minimum entropy production,
\begin{align}
2 \delta(\bm{p}_\tau,\bm{p}_0) \artanh \bigg(&\frac{\delta(\bm{p}_\tau,\bm{p}_0)}{\av{M}} \bigg) \leq \Delta S^{\text{irr},*} \leq 2 d(\mathcal{G}) \artanh \bigg(\frac{d(\mathcal{G})}{\av{M}} \bigg).
\end{align}
The lower bound depends only on the initial and final state via their total variation distance and is independent of the connectivity of the state space.
$\delta(\bm{p}_\tau,\bm{p}_0)$ is the Wasserstein distance on a complete graph.
By contrast, the upper bound depends only on the connectivity of the state space and is independent of the initial and final state.
$d(\mathcal{G})$ equals the Wasserstein distance between an initial and final state each concentrated on one of two maximally distant vertices.

\section{Relation to Langevin dynamics}  \label{sec-benamou}
The results of the preceding section imply three alternative but equivalent ways of formulating the optimal transport problem of computing the Wasserstein distance \eqref{wasserstein}.
All of them involve constructing a Markov jump process of the form \eqref{master} interpolating between the two probability vectors $\bm{q} = \bm{p}_0$ and $\bm{p} = \bm{p}_\tau$.
\begin{enumerate}
\item Find the Markov jump process with initial state $\bm{p}_0 = \bm{q}$ and final state $\bm{p}_\tau = \bm{p}$ that minimizes the average number of transitions $\av{M}$.
Then, the Wasserstein distance is given by $\mathcal{W}(\bm{p},\bm{q}) = \av{M}^*$.
\item Find the Markov jump process with initial state $\bm{p}_0 = \bm{q}$ and final state $\bm{p}_\tau = \bm{p}$ that minimizes the entropy production $\Delta S^\text{irr}$, \eqref{entropy}, at a given average number of transitions $\av{M}$.
Then, the Wasserstein distance is related to the minimum entropy production via \eqref{minent-finite}.
\item Find the Markov jump process with initial state $\bm{p}_0 = \bm{q}$ and final state $\bm{p}_\tau = \bm{p}$ that minimizes the pseudo entropy production $\Delta R^\text{irr} = \int_0^\tau dt \ \rho_t$, \eqref{pseudo-def}, at a given average number of transitions $\av{M}$.
Then, the Wasserstein distance is related to the minimum pseudo entropy production via
\begin{align}
\Delta R^{\text{irr},*} = \frac{2\mathcal{W}(\bm{p},\bm{q})^2}{\av{M}} = \frac{2\mathcal{W}(\bm{p},\bm{q})^2}{\tau \bar{\chi}} \label{pseudo-wasserstein} .
\end{align}
\end{enumerate}
The last formulation, in particular, is appealingly similar to the well-known Benamou-Brenier formula \cite{Ben00} for the $L^2$-Wasserstein distance in $\mathbb{R}^n$.
We recall \eqref{current} defining the probability current $J_t(i,j)$ and traffic $D_t(i,j)$,
\begin{align}
J_t(i,j) =  W_t(i,j) p_t(j) - W_t(j,i) p_t(i) , \qquad D_t(i,j) =  W_t(i,j) p_t(j) + W_t(j,i) p_t(i) .
\end{align}
We define the discrete gradient and divergence operator
\begin{align}
\big(\text{grad}(a)\big)(i,j) &= \frac{1}{2} \big(a(i) - a(j) \big), \\
\big(\text{div}(A) \big)(i) &= \frac{1}{2}  \sum_{j} \big( A(j,i) - A(i,j) \big) \n.
\end{align}
Note that the gradient operator defines an antisymmetric $N \times N$-matrix in terms of the $N$-vector $\bm{a}$, while the divergence operator defines an $N$-vector in terms of the antisymmetric part of the $N \times N$-matrix $A$.
With this notation, we can write \eqref{master} as a continuity equation
\begin{align}
d_t \bm{p}_t = -\text{div}(J_t) \label{continuity-markov}.
\end{align}
Let us further define the inner products for vectors and matrices, respectively
\begin{align}
\Av{\bm{a},\bm{b}} = \sum_{i} a(i) b(i) \quad \text{and} \quad \Av{A,B} = \sum_{i,j} A(i,j) B(i,j) .
\end{align}
With this definition, it is straightforward to check that we have the discrete integration by parts formula
\begin{align}
\Av{\bm{a},\text{div}(A)} = -\Av{\text{grad}(\bm{a}),A} \label{partial-integration}.
\end{align}
The pseudo-entropy production and average number of transitions can be written as
\begin{subequations}
\begin{align}
\Delta R^\text{irr} = \int_0^\tau dt \ \Av{J_t,\frac{J_t}{D_t}} \label{entropy-inner} \\
\av{M} = \frac{1}{2} \int_0^\tau dt \ \av{D_t,1_{N\times N}} \label{transition-traffic} ,
\end{align}
\end{subequations}
where the division is taken as component-wise division and $1_{N\times N}$ is a matrix with all entries equal to $1$.
In summary, the statement of the optimal transport problem in terms of the pseudo-entropy is the following:
Find
\begin{align}
\Delta R^{\text{irr},*} = \inf_{J_t,D_t} \int_0^\tau dt \ \Av{J_t,\frac{J_t}{D_t}} \label{optimization-markov}
\end{align}
such that the continuity equation \eqref{continuity-markov} with boundary conditions $\bm{p}_0 = \bm{q}$ and $\bm{p}_\tau = \bm{p}$ the constraint \eqref{transition-traffic} are satisfied.
Then, the Wasserstein distance is related to the minimum pseudo-entropy production by  \eqref{pseudo-wasserstein}.
By comparison, for a Langevin dynamics ($\bm{x} \in \mathbb{R}^n$) with diffusion coefficient $\mathcal{D}$, the probability density $p_t(\bm{x})$ satisfies the Fokker-Planck equation
\begin{align}
\partial_t p_t(\bm{x}) &= - \text{div}\big(\bm{j}_t(\bm{x})\big) \qquad \text{with} \label{continuity} \qquad
\bm{j}_t(\bm{x}) = \bm{f}_t(\bm{x}) p_t(\bm{x}) - \mathcal{D} \ \text{grad} \big( p_t(\bm{x}) \big)  
\end{align}
and the entropy production is given by
\begin{align}
\Delta S^\text{irr} = \int_0^\tau dt \int d\bm{x} \ \frac{\vert \bm{j}_t(\bm{x}) \vert^2}{\mathcal{D} p_t(\bm{x})} = \int_0^\tau dt \Av{\bm{j}_t,\frac{\bm{j}_t}{\mathcal{D} p_t}} ,
\end{align}
where the inner product is defined as
\begin{align}
\Av{\bm{a},\bm{b}} = \int d\bm{x} \ \bm{a}(\bm{x}) \cdot \bm{b}(\bm{x}),
\end{align}
where $\cdot$ is the standard scalar product in $\mathbb{R}^n$.
As has been discussed in Refs.~\cite{Aur11,Aur12,Dec19b}, we may consider the optimization problem
\begin{align}
\Delta S^{\text{irr},*} = \inf_{\bm{j}_t,p_t} \int_0^\tau dt \ \Av{\bm{j}_t,\frac{\bm{j}_t}{\mathcal{D} p_t}} \label{optimization-langevin}
\end{align}
under the constraint that the continuity equation \eqref{continuity} with boundary conditions $p(\bm{x})_0 = q(\bm{x})$ and $p(\bm{x})_\tau = p(\bm{x})$ is satisfied.
The solution to this problem is related to the Wasserstein distance of order 2 between $p_0(\bm{x})$ and $p_\tau(\bm{x})$ via
\begin{align}
\Delta S^{\text{irr},*} = \frac{\mathcal{W}_2(p_\tau,p_0)^2}{\tau \mathcal{D}} \label{entropy-wasserstein-langevin} .
\end{align}
The latter is defined as
\begin{align}
\mathcal{W}_2(p_\tau,p_0) = \Bigg( \inf_{\Pi} \int d\bm{x} \int d\bm{y} \ \Vert \bm{x} - \bm{y} \Vert^2 \Pi(\bm{x},\bm{y}) \Bigg)^{\frac{1}{2}},
\end{align}
where $\Pi(\bm{x},\bm{y})$ is a joint probability density with marginals $p_\tau(\bm{x})$ and $p_0(\bm{y})$.
The resemblance between \eqref{optimization-markov} and \eqref{optimization-langevin} is obvious.
There are, however, two crucial differences between the two results:
First, the Wasserstein distance in \eqref{entropy-wasserstein-langevin} is defined in terms of the order 2 distance on euclidean space, whereas \eqref{pseudo-wasserstein} involves the order 1 distance on a graph.
And second, the quantity in the denominator is different, specifically, the diffusion coefficient $\mathcal{D}$ appears in the Langevin case and the time-averaged activity $\bar{\chi}$ in the Markov jump case.
However, given their striking similarity, we conjecture that \eqref{entropy-wasserstein-langevin} indeed emerges from \eqref{pseudo-wasserstein}, provided that the jump dynamics \eqref{master} tends to a Langevin dynamics in a suitable continuum limit.
In this limit, we expect that the constraint on the number of transitions is equivalent to fixing the diffusion coefficient $\mathcal{D}$.
We further remark that, in the continuum limit, the quantities $\Delta S^\text{irr}$ and $\Delta R^\text{irr}$ become equivalent, so both problems involve a minimization of the entropy production under constraints.

There is, however, another, more fundamental difference between the continuous and discrete case:
Let us return to the formulation of Section \ref{sec-minent}, that is, we fix the time evolution of the occupation probabilities $\bm{p}_t$ and the activity $\chi_t$ and minimize the entropy production rate $\sigma_t$.
For the Langevin case, a similar setup---fixing the diffusion coefficient instead of the activity---was discussed in Refs.~\cite{Mae14,Dec21}.
In that case, it was found that the force that realizes the minimum entropy production dynamics is the gradient of a potential, which is the equivalent of the detailed balance condition \eqref{db-rates}.
In the Langevin case, there is further a one-to-one correspondence between the potential and the time-dependent probability density, that is, there is only one potential (up to an additive constant) that gives rise to a given time-evolution.
As a consequence, if the dynamics is described by a potential force, this force already minimizes the entropy production rate.
As we found in Section \ref{sec-minent}, the antisymmetric part of the transition rates also satisfies the detailed balance condition and thus is derived from a potential.
However, in contrast to the Langevin case, the one-to-one relation between the potential and the time-evolution of $\bm{p}_t$ is lost.
The reason is that, in principle, we may vary the symmetric and antisymmetric part of the transition rates separately, so that different potentials may give rise to the same time-evolution.
This is clear from the potential landscape of the minimum entropy production dynamics, which has a fixed energy difference between any two connected states, while this condition is not satisfied by a general potential.
This means that, even if the rates of the original dynamics satisfy the detailed balance condition \eqref{db-rates}, we can generally reduce the entropy production rate further while keeping the time-evolution and the activity invariant.
The consequence of this difference is that, while in the Langevin case, the minimum entropy production for a given time-evolution can be identified as the excess entropy production \cite{Mae14} (in the sense of non-equilibrium thermodynamics, see also Refs.~\cite{Oon98,Hat01,Kom08}) and the remaining entropy production as the housekeeping part, this identification is not straightforward in the Markov jump case.


\section{Illustrative examples}  \label{sec-examples}

\subsection{Simple networks}
\begin{figure}
\includegraphics[width=.48\textwidth]{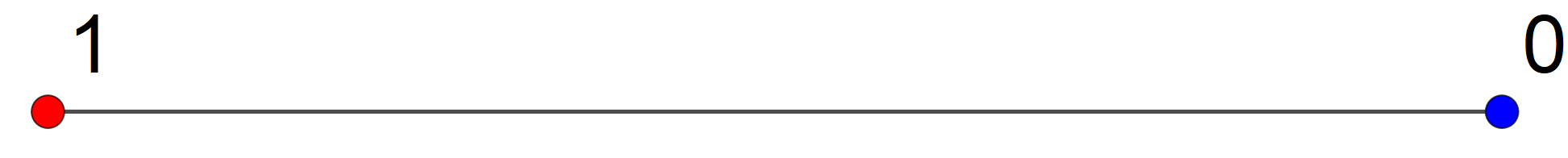}\\
\includegraphics[width=.48\textwidth]{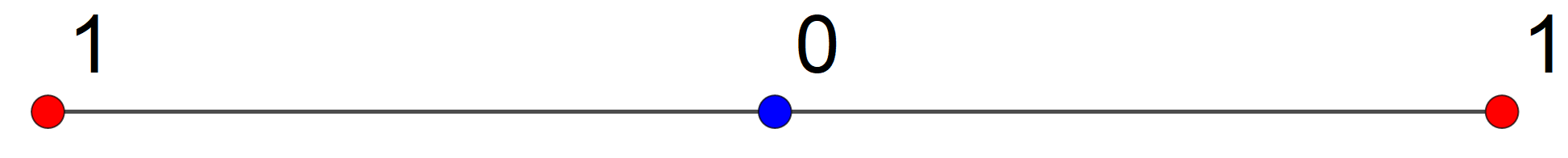}\\
\includegraphics[width=.48\textwidth]{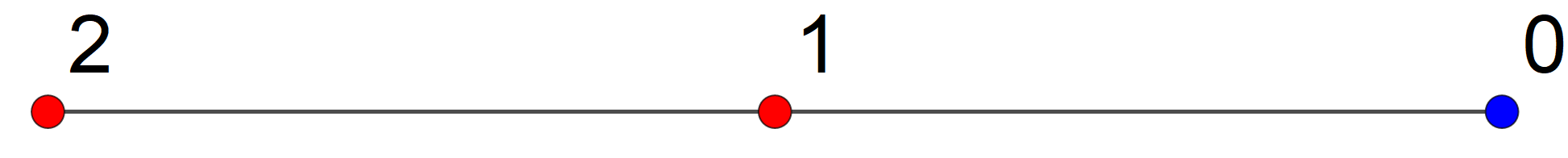}\\
\includegraphics[width=.48\textwidth]{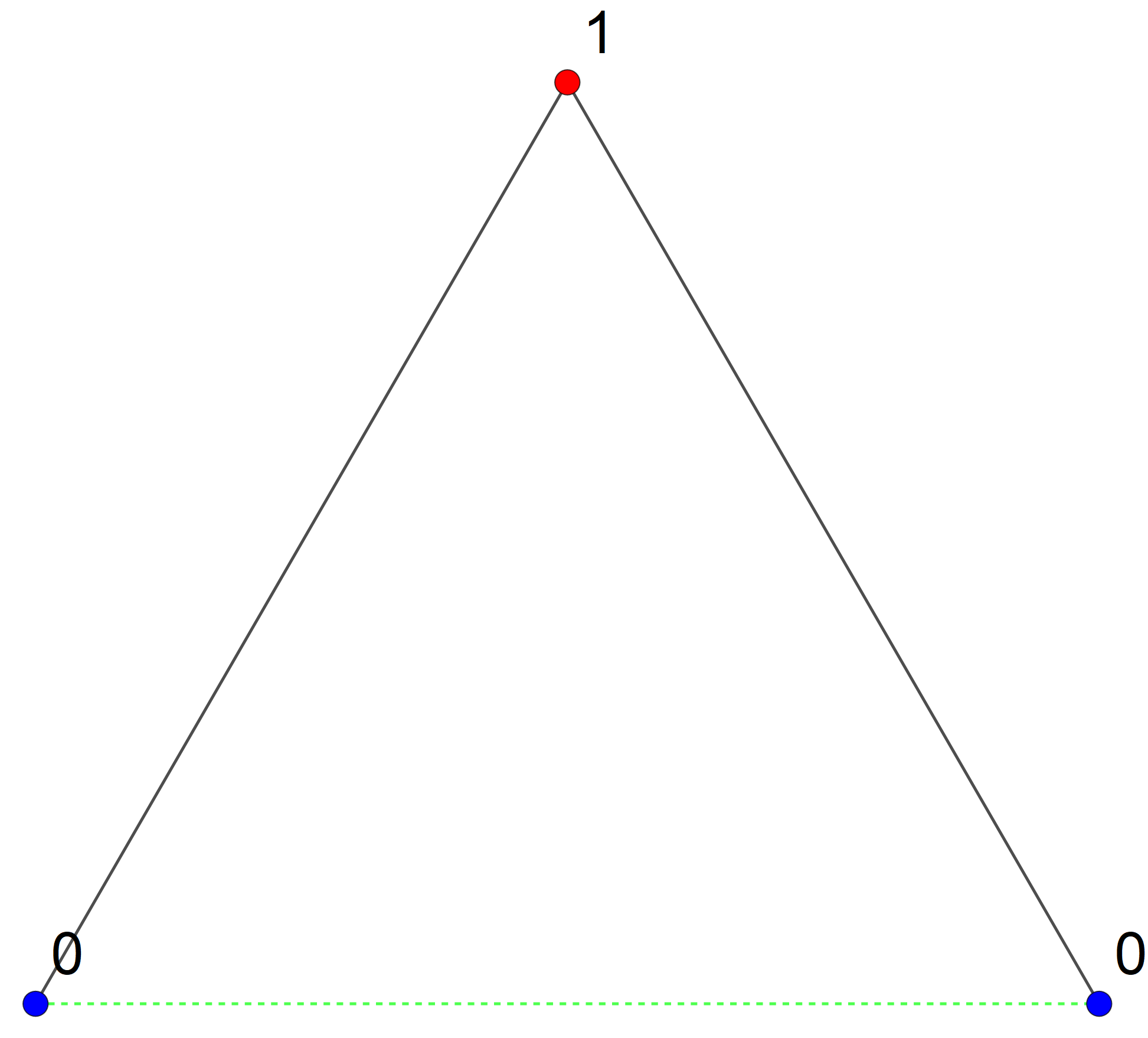}
\caption{Possible configurations for $N=2$ and $N=3$ states. Red dots represent states with $d_t p_t(i) < 0$ (sources of probability), while blue dots represent states with $d_t p_t(i) > 0$ (sinks of probability). The numbers are the energies $\psi(i)$ of each state, chosen such that the lowest energy is zero. Solid lines are connect states with a non-vanishing probability flow between them in the minimum entropy production dynamics, while the green dashed lines connect states between which a connection is permitted in principle ($k(i,j) = 1$), but between which there is no probability flow in the minimum entropy production dynamics. From top to bottom: For $N=2$ states, there is only one possible non-trivial configuration and the Wasserstein distance is equal to the total variation distance. For $N=3$ states on a line, the Wasserstein distance may be equal (upper configuration) or larger (lower configuration) than the total variation distance, depending on the change in the occupation probabilities. Note that the second configuration has a maximal energy difference of $3$. For a loop with $N=3$ states, two states are always effectively disconnected in the minimum entropy production dynamics and the Wasserstein distance is equal to the total variation distance. } \label{fig-3-states}
\end{figure}
\begin{figure}
\includegraphics[width=.48\textwidth]{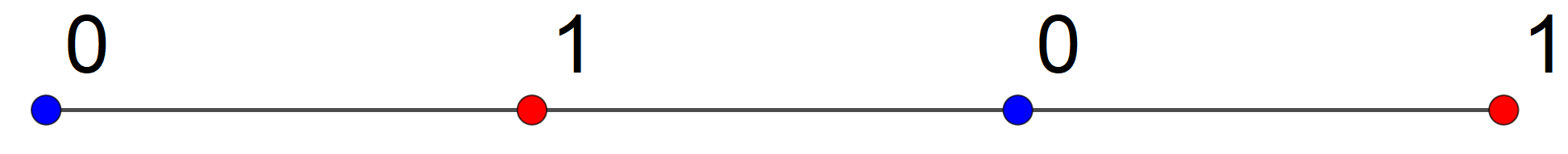}\\
\includegraphics[width=.48\textwidth]{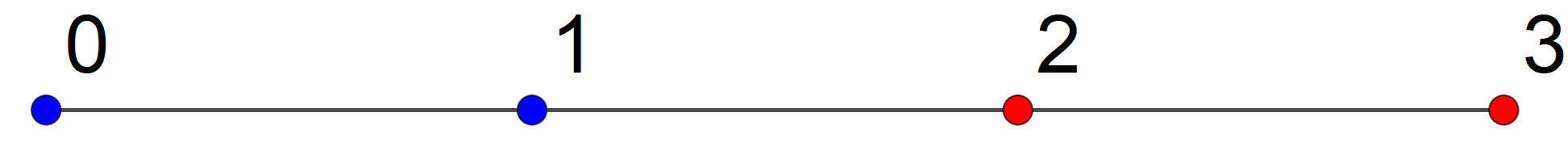}\\
\includegraphics[width=.48\textwidth]{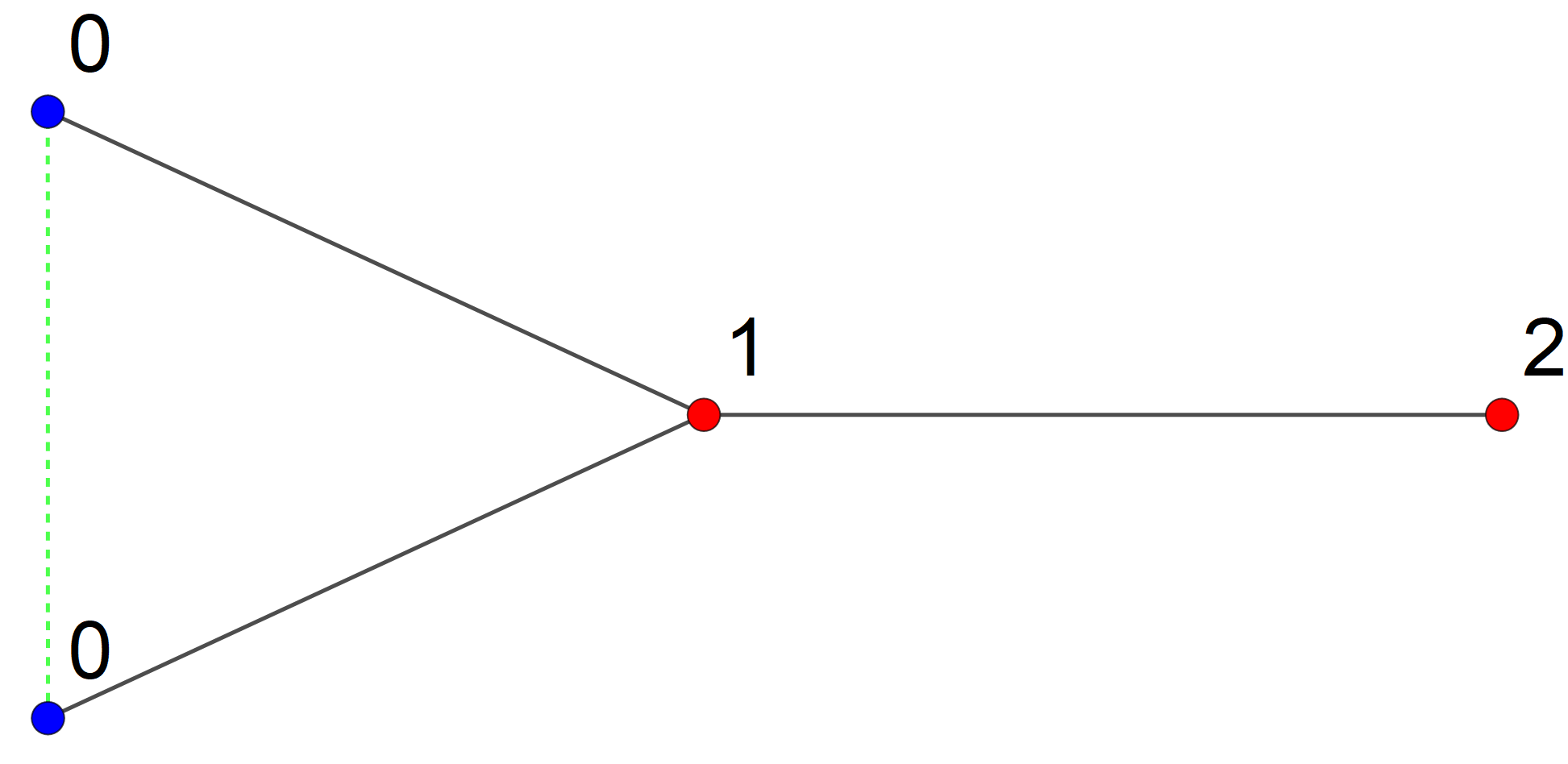}\\
\includegraphics[width=.48\textwidth]{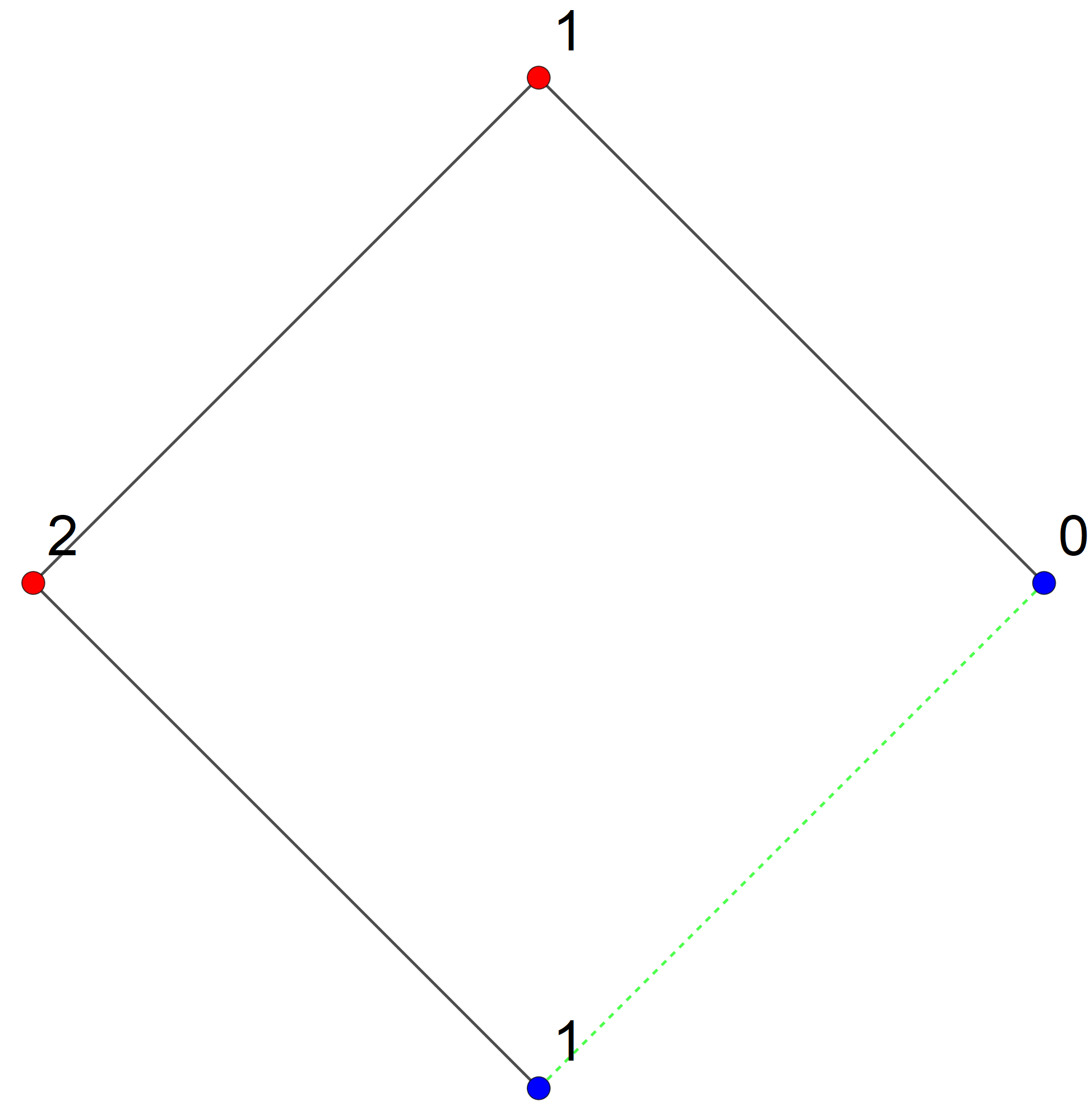}
\caption{Some configurations for $N=4$ states. The symbols are the same as in Fig.~\ref{fig-3-states}. As for three states, a linear configuration may lead to a Wasserstein distance that is equal to (upper configuration) or larger than (lower configuration) the total variation distance. If the original graph contains any odd-numbered loop, those loops are cut in the minimum entropy production dynamics. Finally, for an even-numbered loop, there always exists a tree-like configuration. However, in this case, we may also transport (some or all of the) probability from the left to the right state via the lower path without changing the Wasserstein distance, so we can also find an equivalent configuration which contains the loop. } \label{fig-4-states}
\end{figure}
\begin{figure}
\includegraphics[width=.48\textwidth]{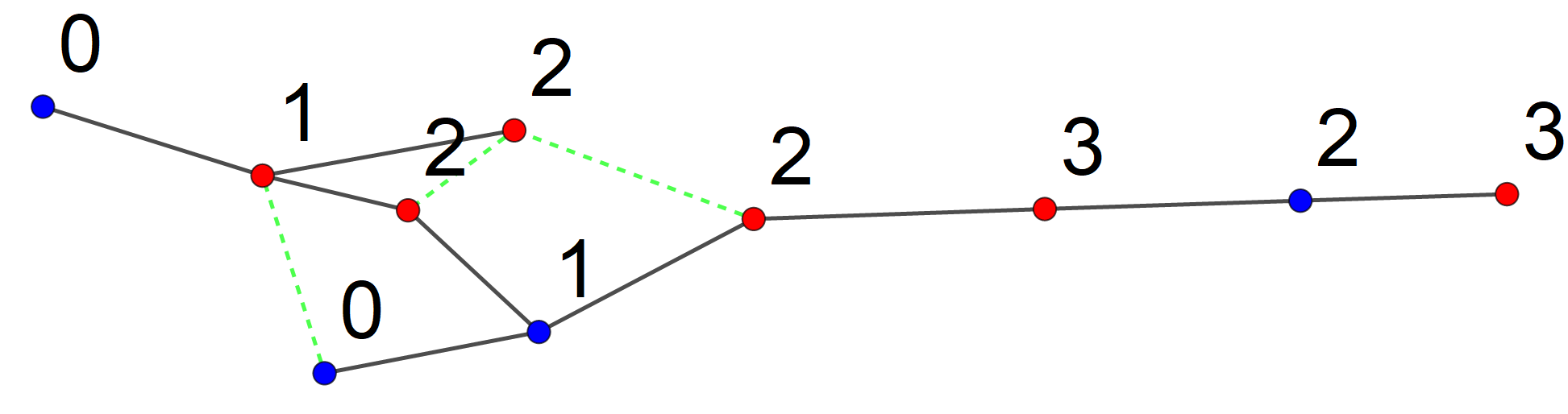}\\
\includegraphics[width=.48\textwidth]{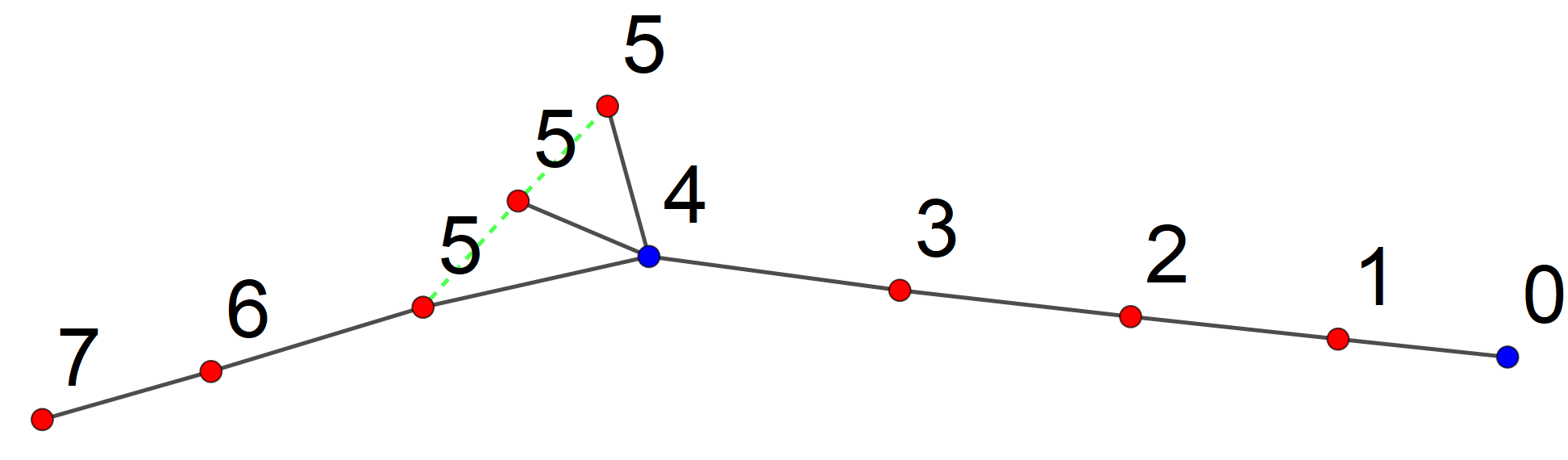}
\caption{Two sample configurations for $N=10$ states. The symbols are the same as in Fig.~\ref{fig-3-states}. For the upper configuration, the Wasserstein distance is about $1.43$ times, for the lower configuration about $3.52$ times the total variation distance.} \label{fig-10-states}
\end{figure}
While the matrix $k(i,j)$ specifies the connectivity of the state network of the original dynamics, for the minimum entropy production dynamics, it is the maximal connectivity.
The difference between the two is that, even if a connection between two states is allowed in principle ($k(i,j) = 1$), the transition rates \eqref{rates-mod-optimal} between the two states may be zero.
Physically, this corresponds to inserting a large energy barrier between the two states, effectively preventing transitions between them.
One particular example for this are loops consisting of an odd number of states.
Along such a loop, it is not possible to define an energy landscape with constant step size in a consistent manner.
Thus, the minimum entropy production dynamics never contains odd loops.
While loops with an even number of states are allowed in principle, we can always find a choice for the optimal coupling $\Pi^*$ that does not contain any loops, such that the corresponding graph is a tree. 
Precisely speaking, we may also have a collection of trees without any connection between them, provided that the total change in probability in any one tree is zero.
The ambiguity in the choice of the optimal coupling reflects the non-uniqueness of the graph distance in the presence of even loops: 
States on opposite sides of the loop are always connected by (at least) two paths of the same length, so we can move probability between them along either path without changing the value of the Wasserstein distance.

Let us consider a few simple state networks.
For $N = 2$ states, we always have one state with $d_t p_t(1) > 0$ and one state with $d_t p_t(2) < 0$.
Then, we assign the energies $\psi(1) = 0$ and $\psi(2) = 1$, which ensures that probability flows from state $2$ to state $1$.
In this case, the Wasserstein distance is equal to the total variation distance.
As remarked before, this is always the case for a complete graph, i.~e.~$k(i,j) = 1$ for all pairs of states.
More generally, the Wasserstein distance is also equal to the total variation distance, if any state $i$ with $d_t p_t(i) > 0$ is connected to all states $j$ with $d_t p_t(j) < 0$ and vice versa.
In this case, the energies $\psi(i)$ can be chosen such that $\psi(i) = 0$ for $d_t p_t(i) > 0$ and $\psi(i) = 1$ for $d_t p_t(j) < 0$, that is, the largest energy difference between any two states is $1$.
For $N = 3$, there are two possible configurations that result in a connected graph: 
Either the three states can be arranged on a line, e.~g.~$k(2,1) = k(3,2) = 1$ and $k(1,3) = 0$.
In this case, the connectivity of the minimum entropy production dynamics is generally the same as the original graph.
However, even for this very simple configuration, the Wasserstein distance is generally larger than the total variation distance:
If $d_t p_t(1) > 0$, $d_t p_t(2) > 0$ and $d_t p_t(3) < 0$, then we cannot avoid moving some probability from state $1$ to state $3$, and thus over a graph distance of $d(1,3) = 2$.
The other possibility is that the three states form a loop, e.~g.~$k(2,1) = k(3,2) = k(1,3) = 1$.
In this case, since the loop contains an odd number of states, the minimum entropy production dynamics will always have zero probability flow across one of the edges and we obtain a configuration that can be arranged on a line.
However, since the original graph is complete, the Wasserstein distance is always equal to the total variation distance in this case.
The possible configurations for $2$ and $3$ states are shown in Fig.~\ref{fig-3-states}.
For $N=4$ there are already several qualitatively different possibilities, taking into account the connectivity and distribution of probability changes.
Some examples are shown in Fig.~\ref{fig-4-states}.
The number of possible configurations increases rapidly with the number of states, two random examples for $N=10$ states are shown in Fig.~\ref{fig-10-states}.

\subsection{Pumping current on a ring}
\begin{figure*}
\includegraphics[width=.48\textwidth]{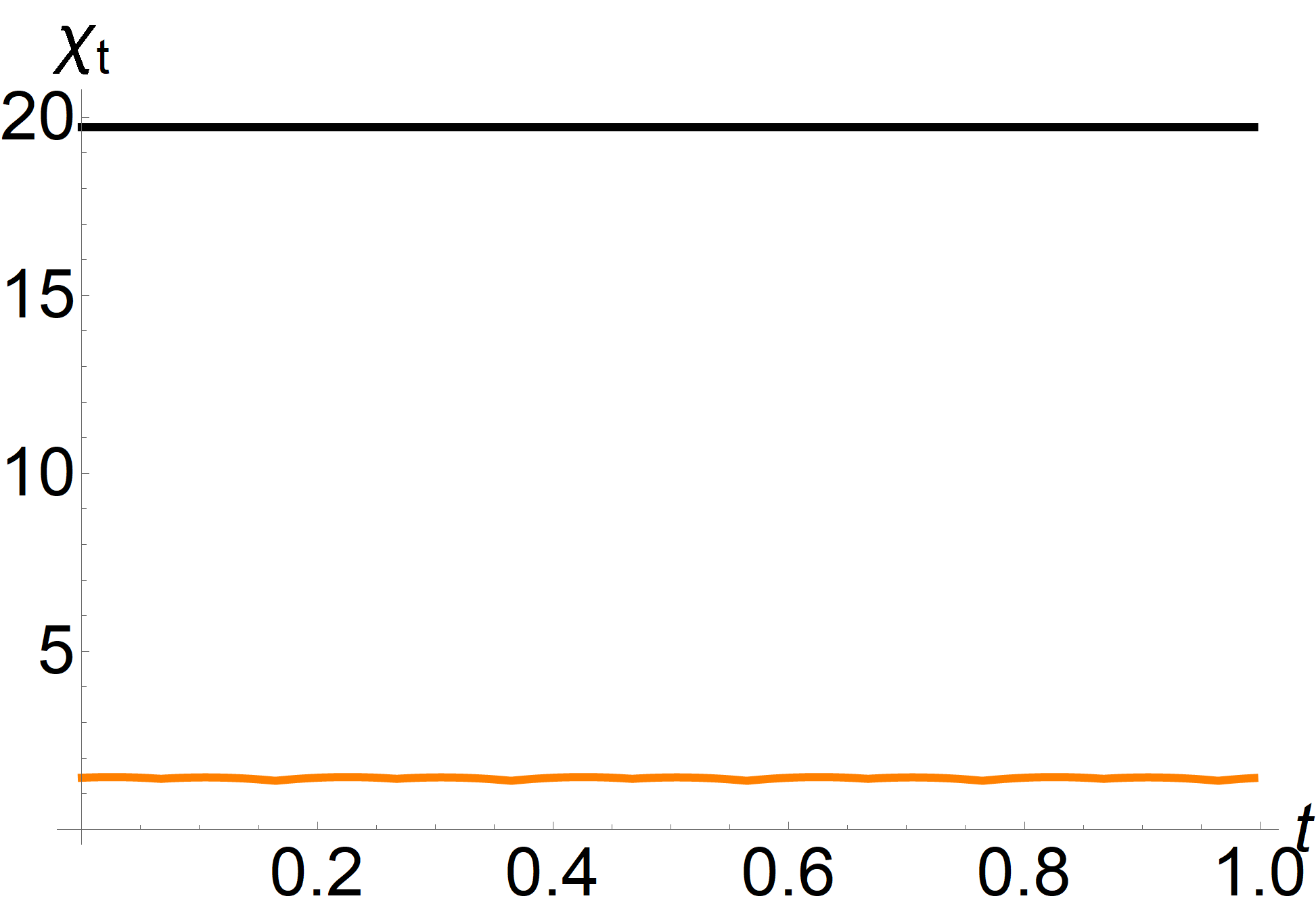}
\includegraphics[width=.48\textwidth]{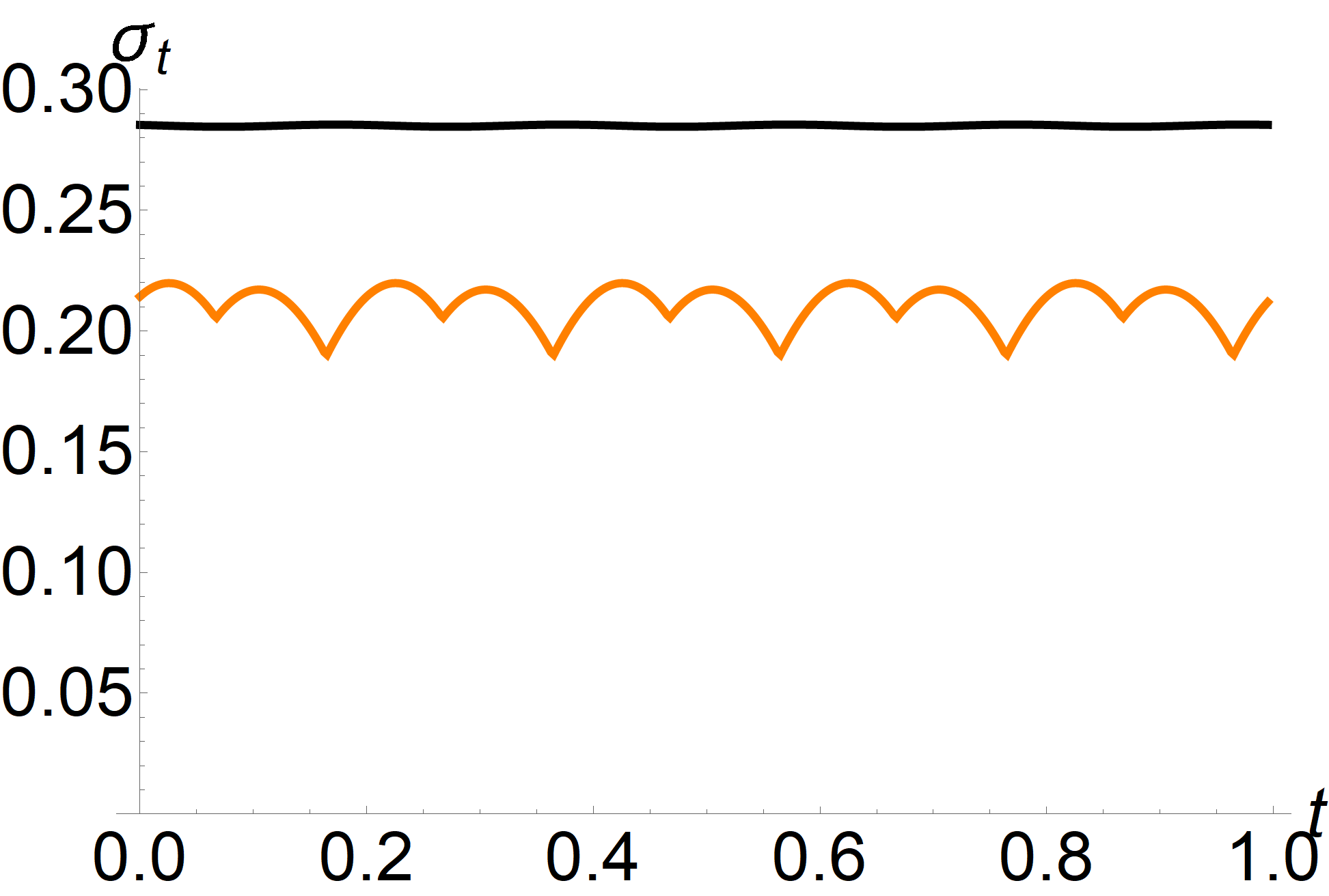}\\
\includegraphics[width=.48\textwidth]{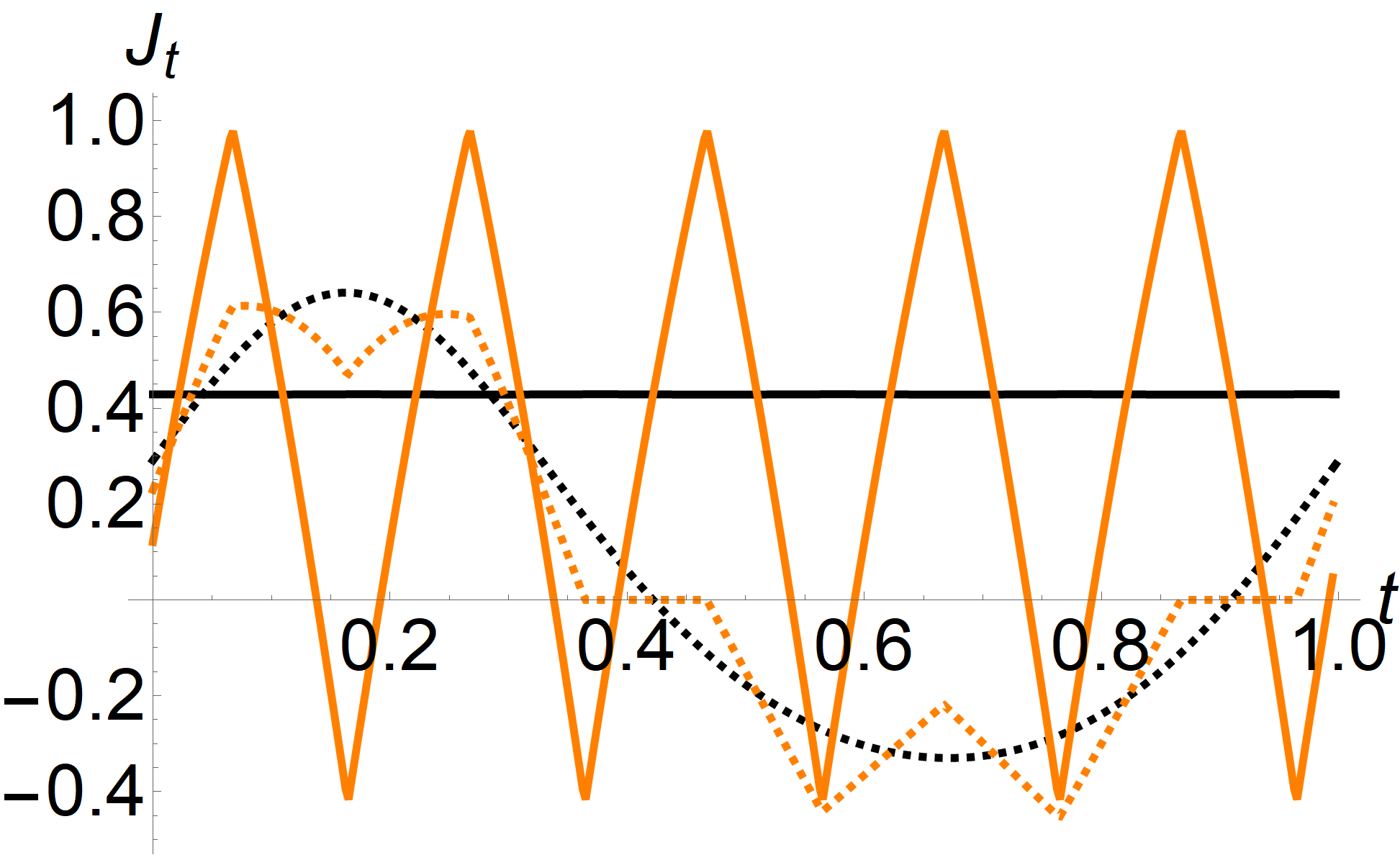}
\includegraphics[width=.48\textwidth]{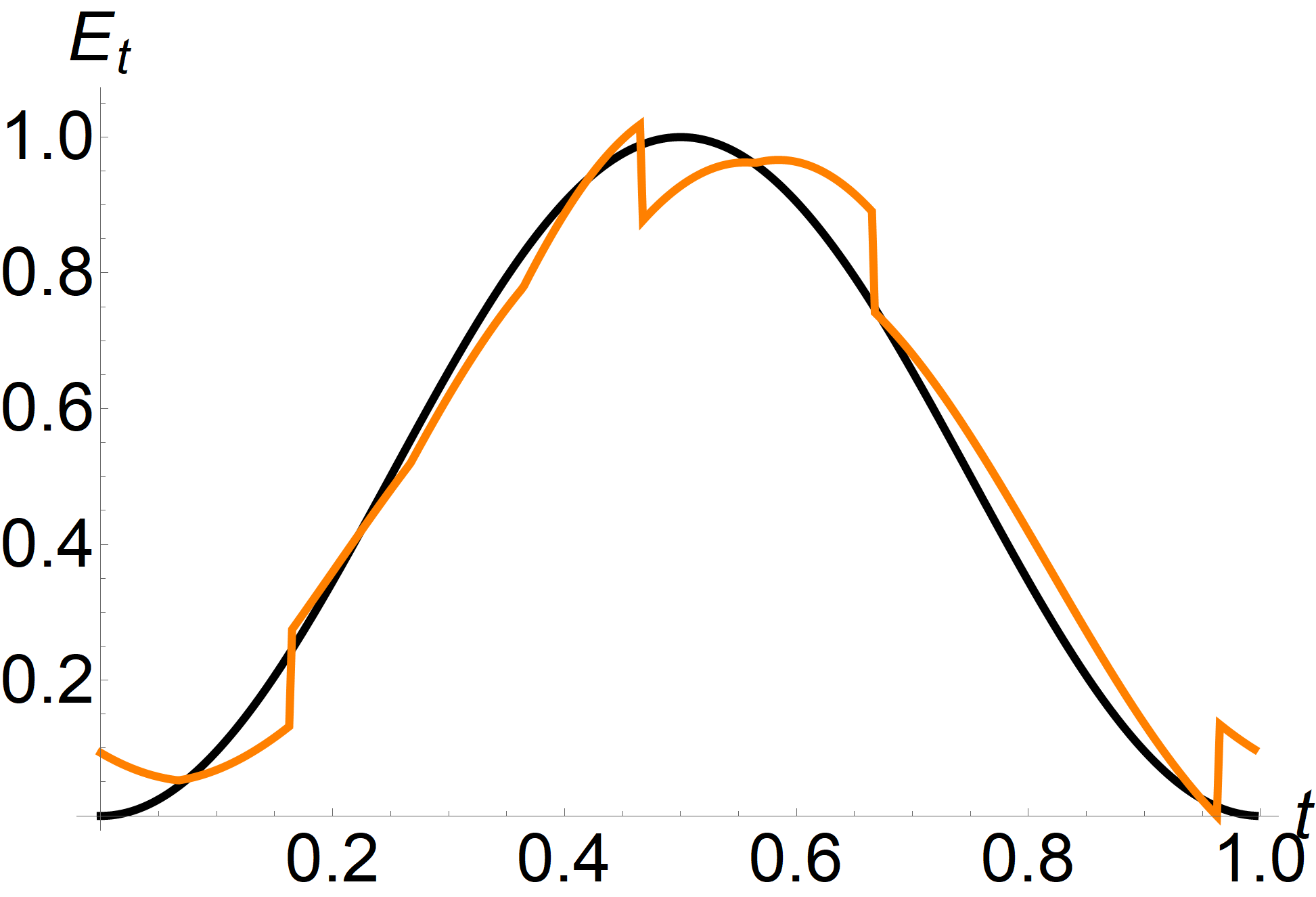}
\caption{Results for the pumping model \eqref{pump-rates} with $N = 5$, $\epsilon_0 = 1$ and $\omega_0 = 10$ as a function of time for one period of the driving. (Top left) The activity $\chi_t$ (black) and the minimal activity $\chi_t^*$ (orange). (Top right) The entropy production rate $\sigma_t$ (black) and the minimum entropy production rate $\sigma_t^*$ (orange). (Bottom left) The probability current $J_t(2,1)$ (dashed) and the total probability current around the ring (solid) for the original dynamics (black) and the minimum entropy production dynamics (orange). (Bottom right) The energy $E_t(1)$ (black, see \eqref{pump-rates}) and the energy $E_t^*(1)$ in the minimum entropy production dynamics (orange, see \eqref{pump-energy}).} \label{fig-pump-fast}
\end{figure*}
\begin{figure*}
\includegraphics[width=.48\textwidth]{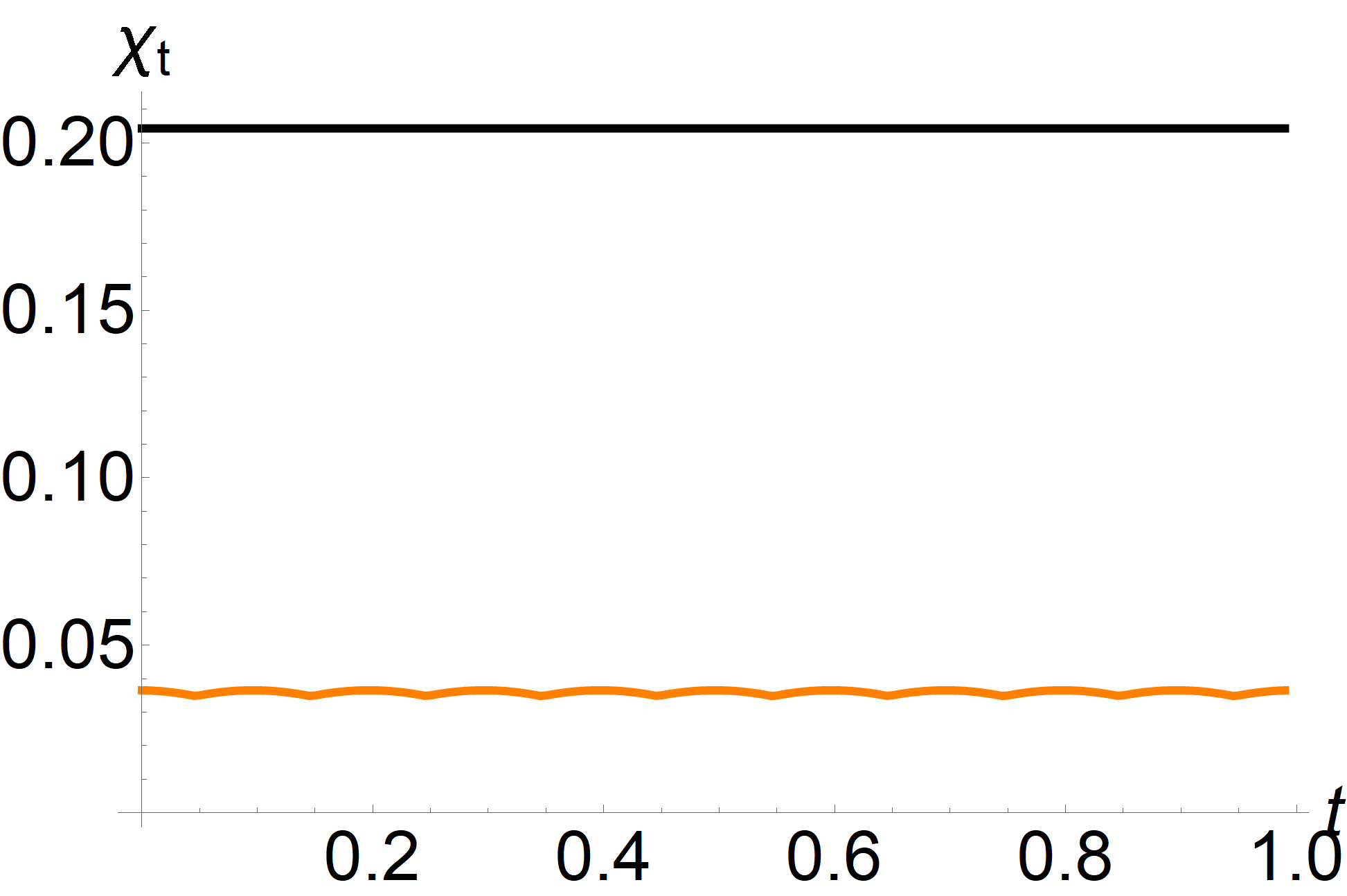}
\includegraphics[width=.48\textwidth]{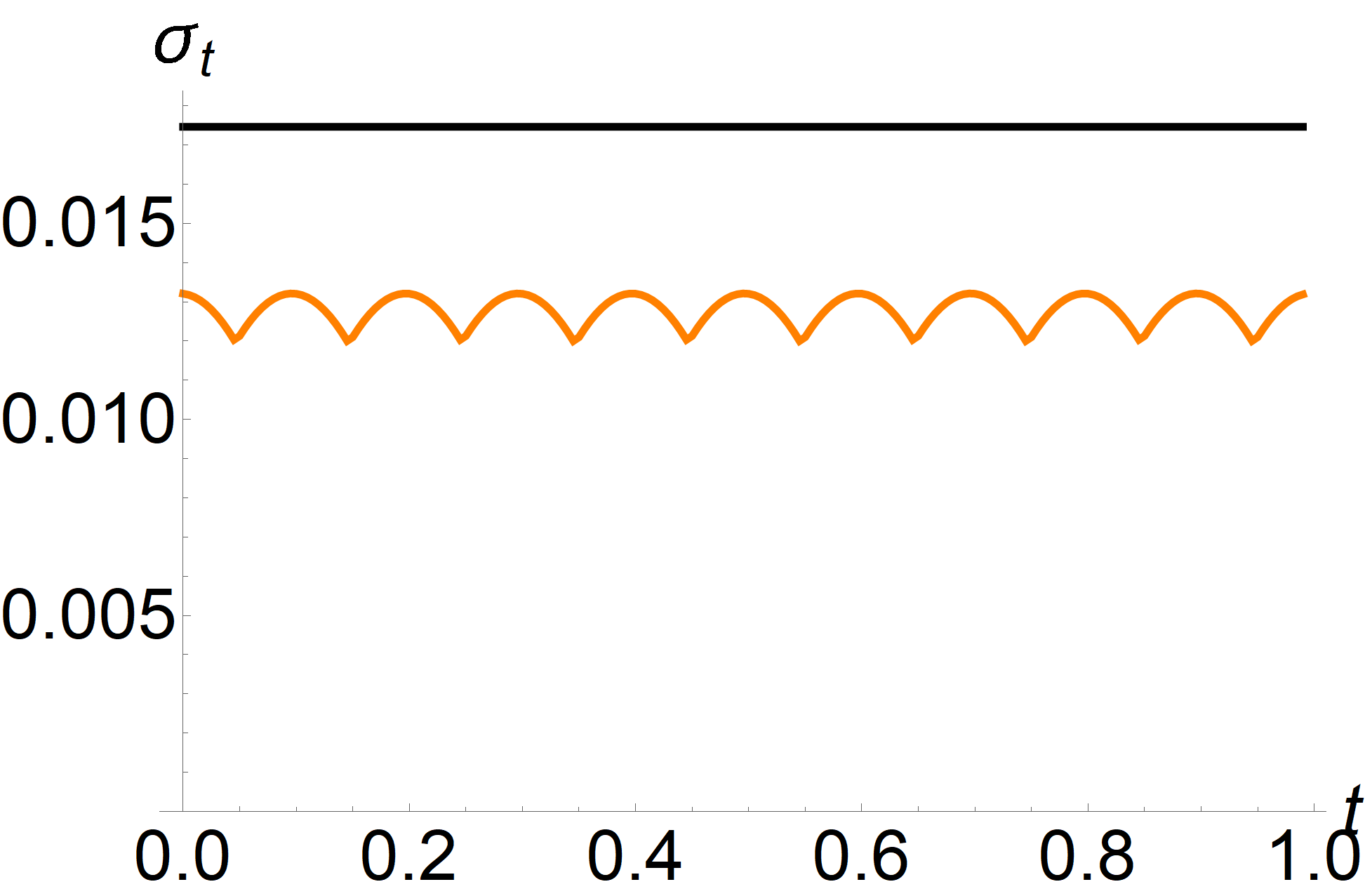}\\
\includegraphics[width=.48\textwidth]{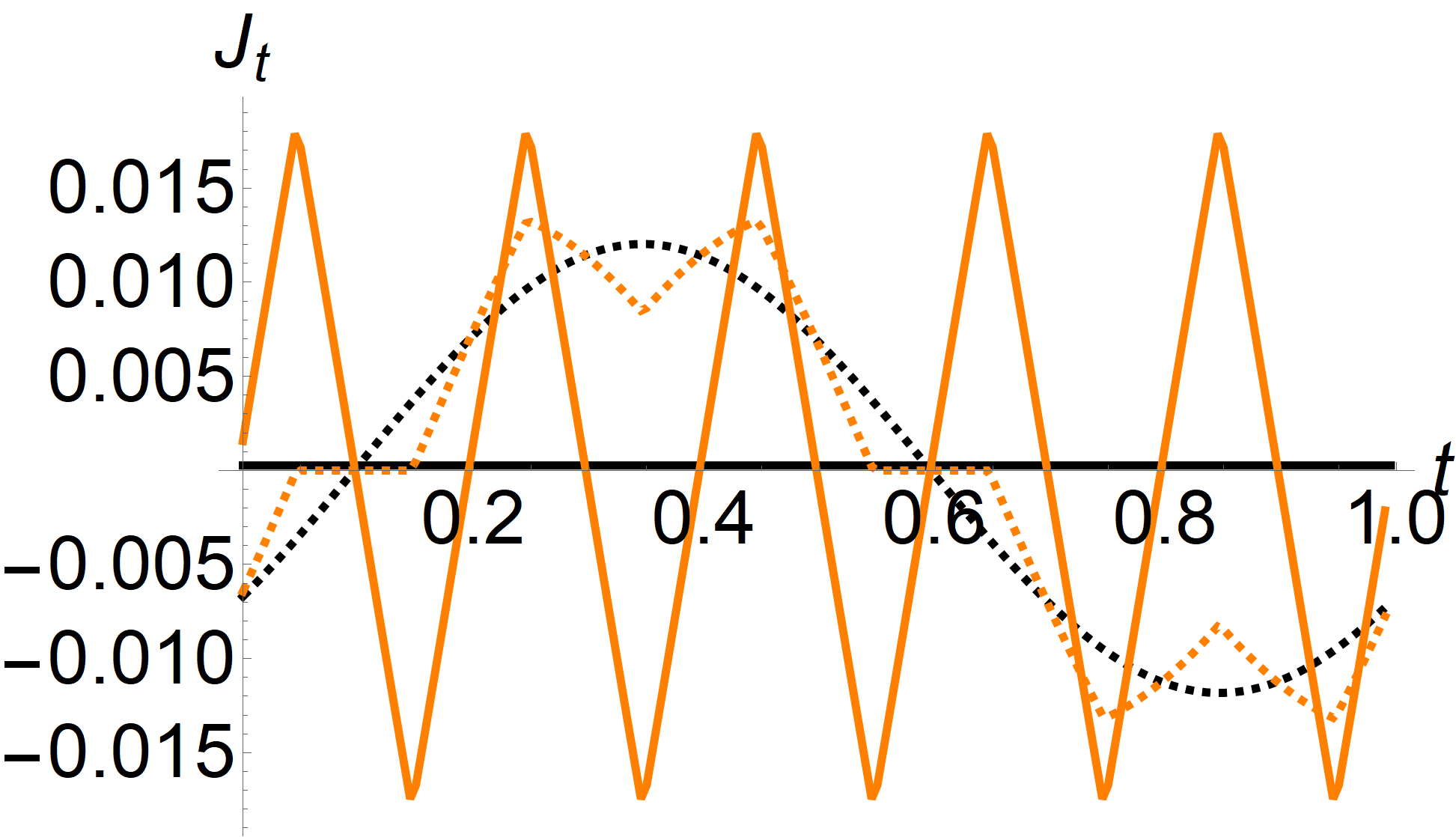}
\includegraphics[width=.48\textwidth]{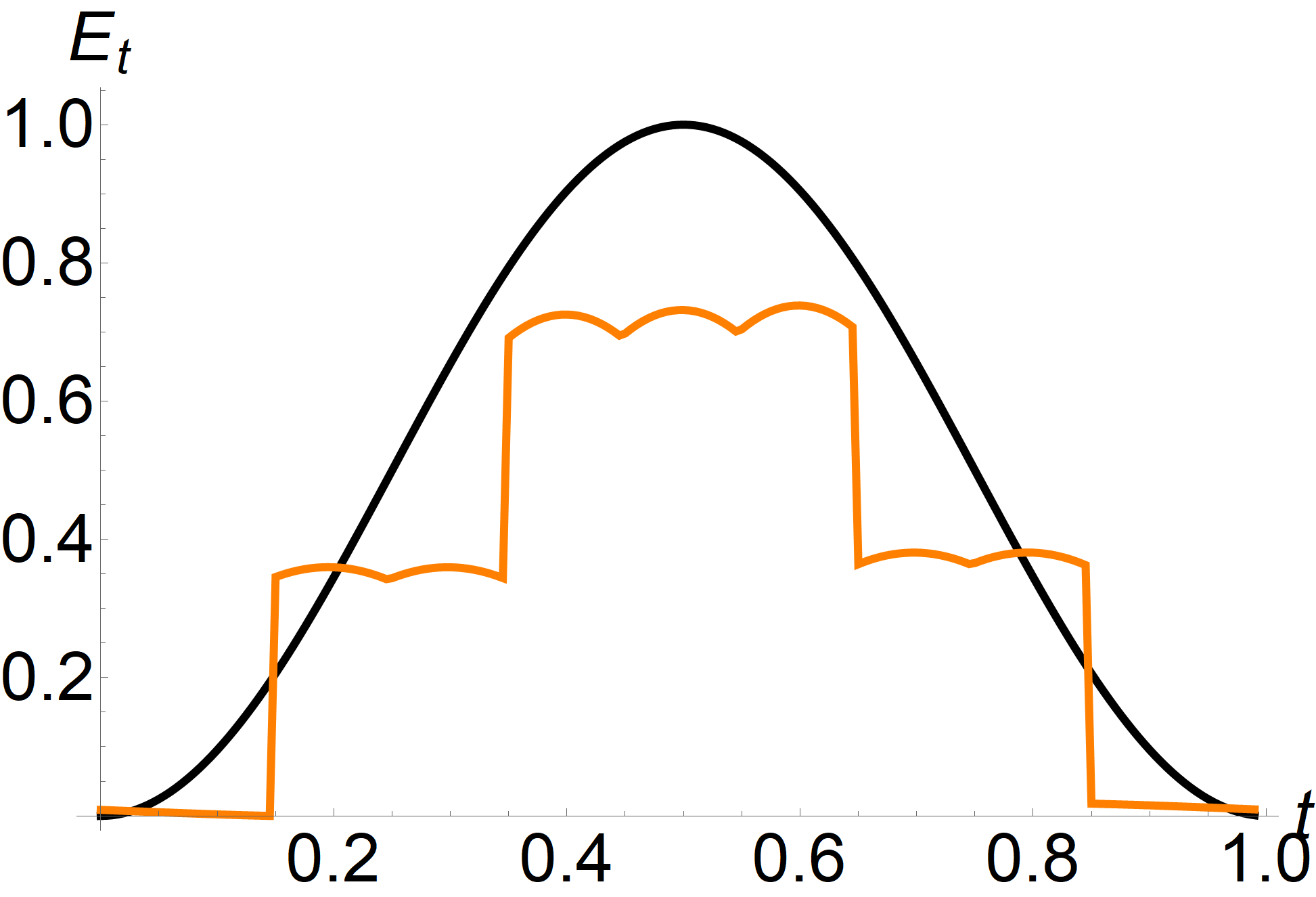}
\caption{Results for the pumping model \eqref{pump-rates} with $N = 5$, $\epsilon_0 = 1$ and $\omega_0 = 0.1$ as a function of time for one period of the driving. (Top left) The activity $\chi_t$ (black) and the minimal activity $\chi_t^*$ (orange). (Top right) The entropy production rate $\sigma_t$ (black) and the minimum entropy production rate $\sigma_t^*$ (orange). (Bottom left) The probability current $J_t(2,1)$ (dashed) and the total probability current around the ring (solid) for the original dynamics (black) and the minimum entropy production dynamics (orange). (Bottom right) The energy $E_t(1)$ (black, see \eqref{pump-rates}) and the energy $E_t^*(1)$ in the minimum entropy production dynamics (orange, see \eqref{pump-energy}).} \label{fig-pump-slow}
\end{figure*}
In the previous section, we only discussed the static structure of the minimum entropy production process for a given initial and final state.
Now we want to extend this to a dynamic situation, where a system is actually driven by an external protocol.
We consider a ring of $N$ states, with transitions between nearest neighbors occurring according to the time-dependent rates
\begin{align}
W_t(i,j) &= \omega_0 \exp \bigg(- \frac{1}{2} \big( E_t(i) - E_t(j) \big) \bigg) \label{pump-rates}   \\
\text{with} \quad E_t(i) &= \epsilon_0 \Bigg(\frac{1}{2} - \frac{1}{2} \cos\bigg(2 \pi \Big(t + \frac{j-1}{N} \Big) \bigg) \Bigg) \n .
\end{align}
The energy of each state varies periodically between $0$ and $\epsilon_0$, with a phase-shift of $2\pi/N$ between two neighboring states.
As a consequence, the location of the minimum energy state moves around the ring and thus the rates \eqref{pump-rates} correspond to pumping a probability current around the ring.
Note that the rates satisfy the detailed balance condition \eqref{db-rates} for any time $t$.
The parameter $\omega_0$ quantifies the rate at which transitions occur between the different states and thus determines how well the system can follow the time-dependence of the rates.
For $\omega_0 \gg 1$, the transitions between the states are fast compared to the change in the rates and thus the occupation probabilities are close to the instantaneous equilibrium probabilities, $p_t(i) \simeq p_t^\text{eq}(i)$ with
\begin{align}
p_t^\text{eq}(i) = \frac{e^{-E_t(i)}}{\sum_j e^{-E_t(j)}} \label{pump-equilibrium} .
\end{align}
For $\omega_0 \ll 1$, on the other hand, the rates change faster than transitions between the individual states occur; in this case, the transitions are approximately determined by the rates averaged over one period and we have $p_t(i) \approx 1/N$.
In the long-time limit, the dynamics \eqref{pump-rates} lead to time-periodic occupation probabilities.
While these do not have an analytic expression, they are easily obtained numerically, which allows us to calculate the entropy production rate $\sigma_t$ \eqref{entropy} and activity $\chi_t$ \eqref{activity} at any given time.
With the time-evolution of the occupation probabilities and the activity, we can then solve the optimization problem in \eqref{c-equations} for any instant of time and thus compute the minimum activity $\chi^*_t$, and the minimum entropy production rate $\sigma_t^*$ at the given activity $\chi_t$.
We can also compute the energy landscape for the minimum entropy production dynamics, which is defined using \eqref{rates-mod-optimal-split} as
\begin{align}
E^*_t(i) = - \ln \big(p_t(i) \big) + 2 \artanh \bigg(\frac{\chi_t^*}{\chi_t} \bigg) \psi_t(i) \label{pump-energy} ,
\end{align}
where $\psi_t(i)$ is determined by the optimal coupling via \eqref{coupling-landscape}.
The results are shown in Fig.~\ref{fig-pump-fast} for fast transitions ($\omega_0 = 10$) in Fig.~\ref{fig-pump-fast} and for slow transitions ($\omega_0 = 0.1$) in Fig.~\ref{fig-pump-slow}.
First, we note that the activity is approximately constant as a function of time.
This means that in this case, we would obtain the same results by fixing the time-averaged activity instead of its instantaneous value.
Next, even though the rates \eqref{pump-rates} already satisfy detailed balance, we can further reduce the entropy production rate while keeping the time-evolution of the occupation probabilities and the activity invariant.
The relative reduction $\sigma_t^*/\sigma_t \approx 0.7$ has the same order of magnitude for both fast and slow transitions.
If the transitions between states are fast compared to the driving, then we observe a finite current and approximately constant around the ring, while for slow transitions, where the system cannot follow the driving, the overall current vanishes.
While this behavior is the same for the current in the minimum entropy production dynamics when averaged over one period, the instantaneous current shows marked oscillations, both for fast and slow transitions.
At first glance, this is counter-intuitive, as it indicates that driving a current back and forth across the ring can reduce the entropy production rate.
However, the latter is actually determined by the local currents between individual states, rather than the global current in the entire ring.
Still, it is not trivial that optimizing the local currents would lead to a more erratic behavior in the overall current.
Finally, with respect to the energy landscape, for fast transitions, we observe that the energy landscape of the minimum entropy production dynamics, \eqref{pump-rates}, closely follows the energies in the original dynamics.
The reason is that, in this limit, the second term in \eqref{pump-energy} is small, since the activity is much larger than the minimum value, so the energy $E_t^*$ is mostly determined by the equilibrium probabilities \eqref{pump-equilibrium} and thus the original energy landscape.
By contrast, for slow transitions, the occupation probabilities are approximately constant and equal and thus enter \eqref{pump-energy} only as a trivial constant shift.
In this limit, we observe a marked step-like behavior of the minimum entropy production energy landscape, which reflects the fact that the direction of the probability flow and thus the optimal coupling change in a non-smooth manner during the time-evolution.

\subsection{Spin flips in the Ising model}
\begin{figure}
\includegraphics[width=.48\textwidth]{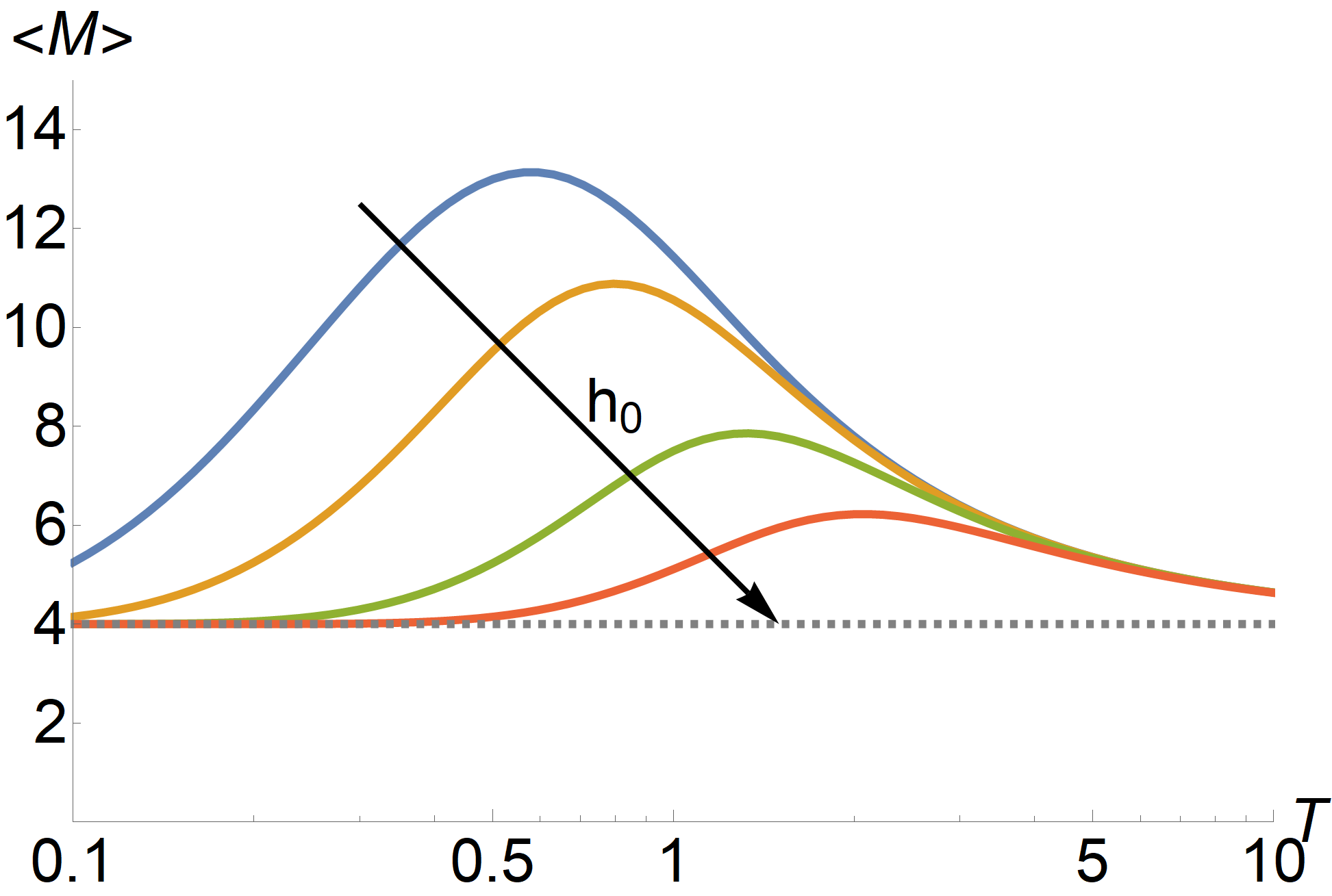}
\includegraphics[width=.48\textwidth]{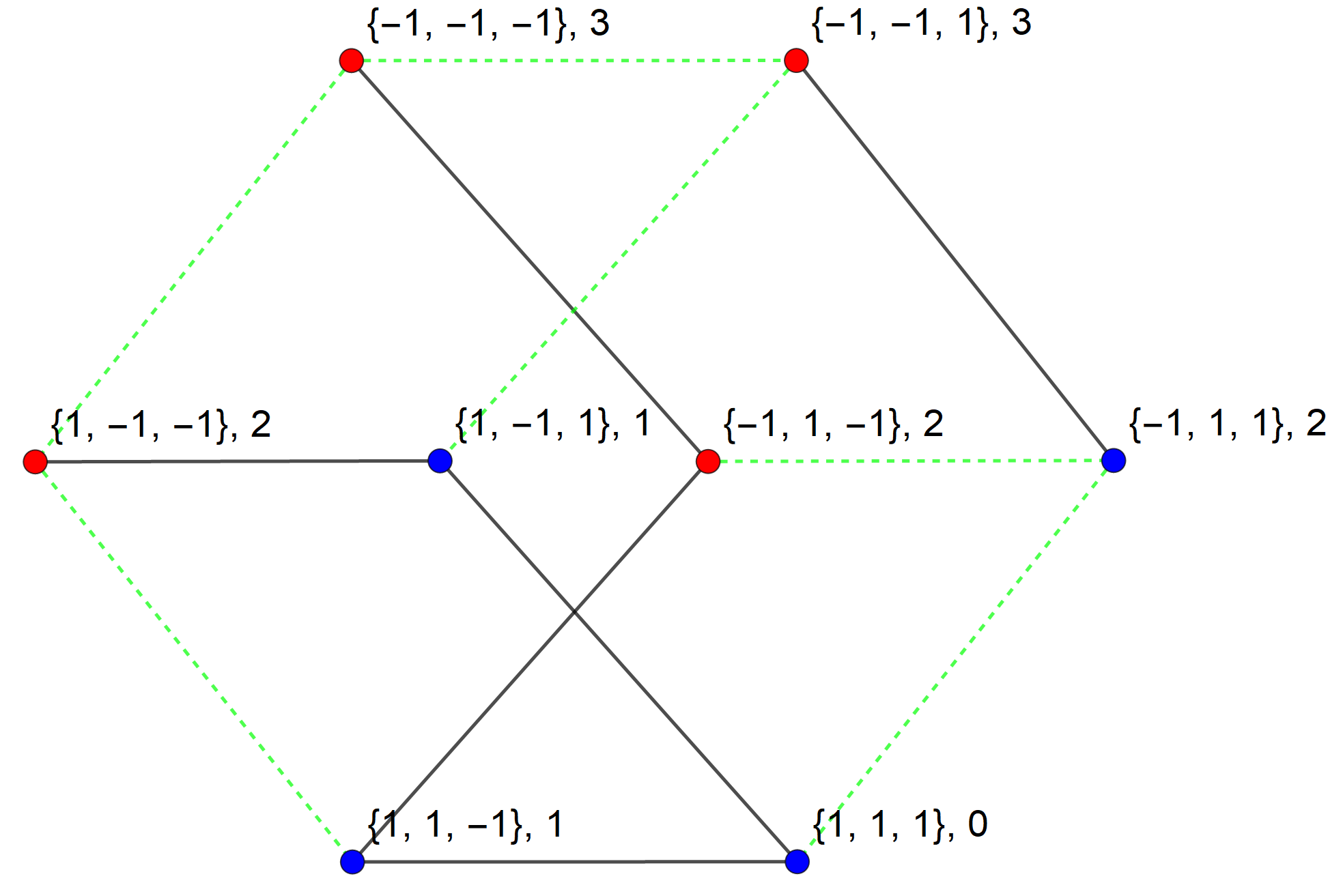}
\caption{(Left panel) The lower bound \eqref{spin-flip-bound} on the number of spin flips for a field reversal in a four-state Ising chain as a function of temperature. The lines correspond to $h_0 = 0.1, 0.2, 0.5, 1.0$ from top to bottom.
(Right panel) The graph corresponding to the minimum entropy production dynamics for a three-state Ising chain under field reversal for $h_0 = 0.5$ and $T = 1.0$. The numbers in braces represent the spin configuration, the second number is the energy $\psi(i)$ of the state; the symbols are the same as in Fig.~\ref{fig-3-states}.} \label{fig-ising}
\end{figure}

The Ising model is a paradigmatic example for a physical system on a discrete state space.
We consider a collection of $K$ classical spins arranged on a $d$-dimensional square lattice.
Each spin can have one of two directions, up (denoted by $\sigma(k) = +1$) and down (denoted by $\sigma(k) = -1$).
The spins interact via a pairwise ferromagnetic interaction between nearest neighbors, characterized by the Hamiltonian
\begin{align}
H_t(\bm{\sigma}) = - J_t \sum_{\av{k l}} \sigma(k) \sigma(l) - h_t \sum_k \sigma(k) \label{ising-hamiltonian} ,
\end{align}
where $J_t$ quantifies the strength of the interaction, $h_t$ is an external field and $\av{k l}$ denotes the sum over nearest-neighbor pairs.
Thus, the first term is minimized when all spins are parallel, whereas the second term is minimized when all spins are aligned with the external field.
For simplicity, we assume that the parameters $J_t$ and $h_t$ are the same for all spins, but may change as a function of time.
The state $i$ of the system is determined by the value of each spin, that is, the vector $\bm{\sigma} = (\sigma(1),\ldots,\sigma(K))$, so that there are $N = 2^K$ different states.
We assume that transitions between different states can occur only via single spin flips, so that $k(i,j) = 1$ if and only if $i$ and $j$ differ by the orientation of exactly one spin, and that the transition rates have the form
\begin{align}
W_t(i,j) =  k(i,j) \omega \exp\bigg({-\frac{\beta_t}{2}\big(H_t(i) - H_t(j) \big)} \bigg) \label{ising-rates} ,
\end{align}
where $\beta_t = 1/T_t$ is the inverse temperature and the constant $\omega$ quantifies the overall speed of the dynamics.
Since the transition rates satisfy detailed balance, the steady state occupation probability (at fixed parameters) is the Boltzmann-Gibbs equilibrium
\begin{align}
p_t^\text{eq}(i) = \frac{e^{-\beta_t H_t(i)}}{\mathcal{Z}_t} \label{ising-equilibrium},
\end{align}
where $\mathcal{Z}_t = \sum_i e^{-\beta_t H_t(i)}$ is the partition function.
The system can be driven out of equilibrium by varying the parameters, typically the external field $h_t$ and temperature $\beta_t$, as a function of time according to a given protocol.
This leads to a time-dependent occupation probability $p_t(i)$ which is generally different from the instantaneous Boltzmann-Gibbs state.
A typical example is a quench, where the system starts out in the equilibrium state corresponding to initial parameters values $h_\text{i}$ and $\beta_\text{i}$.
Then, at $t=0$, the value of the parameters is instantaneously changed to $h_\text{f}$ and $\beta_\text{f}$ and we can observe the random evolution of the system's state and, by repeating this procedure sufficiently often, the occupation probability $p_t(i)$, as it relaxes to the new equilibrium state.
Using \eqref{entropy-medium-shannon}, we can write the total entropy production during the process as
\begin{subequations}
\begin{align}
\Delta S^\text{irr} &= \Delta S^\text{m} + \Delta S^\text{s}  \quad \text{with} \\
\Delta S^\text{m} &= -\int_0^\tau dt \ \beta_t \sum_{i} H_t(i) \partial_t p_t(i) \\
\Delta S^\text{s} &= -\sum_i \Big( \ln \big(p_\tau(i) \big) p_\tau(i) - \ln \big(p_0(i) \big) p_0(i) \Big) .
\end{align}
\end{subequations}
For the case of a quench the value of the temperature and the functional form of the Hamiltonian are independent of time during the process and we can further write
\begin{align}
\Delta S^\text{m} &= - \beta_\text{f} \sum_i H_\text{f}(i) \big( p_\tau(i) - p_0(i) \big) .
\end{align}
The advantage of this formulation is that it only depends on the initial and final occupation probabilities and the Hamiltonian after the quench.
Let us suppose that we wait for a sufficiently long time after the quench so that the occupation probabilities have relaxed to their equilibrium values \eqref{ising-equilibrium}.
Then, the total entropy production is a function of only the initial and final Hamiltonian
\begin{align}
\Delta S^\text{irr} = -\beta_\text{f} \av{H_\text{f}}_\text{i} + \beta_\text{i} \av{H_\text{i}}_\text{i} + \ln \bigg( \frac{\mathcal{Z}_\text{f}}{\mathcal{Z}_\text{i}} \bigg) ,
\end{align}
where $\av{\ldots}_\text{i}$ denotes an average with respect to the Boltzmann-Gibbs equilibrium corresponding to the initial Hamiltonian.
For the graph defined by the transition rates \eqref{ising-rates}, the graph distance between any two states is the number of spins whose direction differs between the two states.
Further, the number of transitions $\av{M}$ is the average number of spin flips during the process and the Wasserstein distance $\mathcal{W}(\bm{p}^\text{eq}_\text{f},\bm{p}^\text{eq}_\text{i})$ corresponds to the minimum average number of spin-flips that is required to transform the initial occupation probability into the final one.
Since the Wasserstein distance likewise only depends on the initial and final occupation probability, we can thus derive a criterion for the number of spin flips during the quench,
\begin{align}
\av{M} \geq \frac{\mathcal{W}(\bm{p}^\text{eq}_\text{f},\bm{p}^\text{eq}_\text{i})}{ \tanh \Big( \frac{\Delta S^\text{irr}}{2 \mathcal{W}(\bm{p}^\text{eq}_\text{f},\bm{p}^\text{eq}_\text{i})} \Big)} \label{spin-flip-bound} .
\end{align}
Note that, unlike the entropy production and the Wasserstein distance, $\av{M}$ depends not only on the initial and final state but on the dynamics of the system after the quench.
Thus, this inequality provides a constraint on the dynamics of the system after a quench.
Note that, so far, we have not used the specific form of the Ising-Hamiltonian \eqref{ising-hamiltonian}, so the above result is valid for any quench dynamics between two equilibrium configurations, even though the interpretation of $\av{M}$ of course depends on the physical system.
As a concrete example, we consider a one-dimensional chain consisting of $K = 4$ spins.
We fix the energy scale by setting $J = 1$ and consider a quench from $h_\text{i} = -h_0$ to $h_\text{f} = h_0$, i.~e.~a reversal of the external field.
The lower bound on $\av{M}$ is shown as a function of the temperature for various values of $h_0$ in Fig.~\ref{fig-ising}.
We see that the number of spin flips is always bounded from below by the number of spins.
For low temperature, this result is expected, since the statistics are dominated by states in which most spins are aligned with the external field, so, on average, we have to flip every spin to reach the new equilibrium configuration.
However, for high temperature, this bound is not so easy to understand, since the contribution from all states is of the same order and the initial and final equilibrium configuration are only slightly different.
Moreover, we find that the bound on the number of spin flips is larger for a weaker external field.
For a strong external field, the spins will align with the external field in an almost deterministic manner; for a weak field, by contrast, the dynamics of the spins is dominated by the interactions between the spins and thus events where a spin randomly aligns with the external field just to flip back because of the orientation of its neighbors increase the number of spin flips during the quench.

Since the number of states grows exponentially with the number of spins, it quickly becomes unfeasible to explicitly compute the Wasserstein distance.
However, as noted above, for low temperature, the statistics is dominated by the states with the minimum energy, which corresponds to fully ordered states in the presence of an external field.
Then, reversing the direction of the field requires us to flip every spin, which means that the Wasserstein distance is approximately equal to $K$.
We thus obtain the simpler approximate bound in the low temperature limit
\begin{align}
\av{M} \geq \frac{K}{ \tanh \Big( \frac{\Delta S^\text{irr}}{2 K} \Big)} \label{precision-spin} .
\end{align}
As noted in \eqref{precision-tradeoff}, this can be interpreted as a tradeoff between precision and dissipation: Reducing the number of spin flips requires increasing the dissipation.
While such a relation is intuitively expected, \eqref{precision-spin} puts it on a quantitative footing.

\section{Discussion}

In this work, we have explored the connection between minimum entropy production, detailed balance and Wasserstein distance. 
Our results suggest several possible new research directions. 
First, as remarked in Section~\ref{sec-benamou}, there is a striking resemblance between the discrete jump and continuous diffusion cases. 
While we conjectured that the two results are indeed equivalent in the continuum limit, this should be shown explicitly. 
The most obvious candidate are hypercubic lattice models that correspond to discrete representations of $\mathbb{R}^d$. 
However, it would be particularly interesting to explore other cases as well, for example triangulations of smooth surfaces.

Second, as remarked in the introduction, another approach to relate entropy production in jump processes to a modified Wasserstein distance has been developed recently \cite{Van21}. 
While technically different from our approach, it would be interesting to see whether the two approaches can be related. 
Further, in Ref.~\cite{Van21}, the same approach was generalized to open quantum systems, so a natural question is whether the current approach also can be applied to this setting. 
The main challenge here is to develop a reasonable graph-based representation of the density matrix as opposed to the probability vector.

Finally, in the Langevin case, the concept of minimum entropy production was used in Ref.~\cite{Mae14} to define a decomposition of the entropy production into excess and housekeeping parts. 
The excess entropy production rate, which is equal to the minimal entropy production rate, vanishes only in a steady state, while the housekeeping part vanishes for a conservative force. 
This allows to separate the dissipation in the system into contributions due to time-dependent driving and non-conservative forces. 
Further, the excess entropy can be expressed in terms of the Wasserstein distance \cite{Dec21}. 
Formally, we may also define the excess part as the minimum  production rate in the Markov jump case and express it in terms of the Wasserstein distance. 
It is tempting to identify the remainder as the housekeeping part. 
However, as we saw in Section~\ref{sec-examples}, the entropy production rate is generally not minimized for an arbitrary conservative force, which implies that the housekeeping part can be non-zero even for conservative forces. 
Thus, it needs to be clarified what the physical meaning of this \enquote{housekeeping entropy} is.

\begin{acknowledgements}
The author gratefully acknowledges inspiring and helpful discussions with S.~Ito and S.-i.~Sasa.
\end{acknowledgements}

\appendix

\section{Proof of \eqref{entropy-bound-2}} \label{app-bound}
\eqref{entropy-bound-2} provides a lower bound on the entropy production rate for a specific choice of the parameter $\mathcal{A}$ in \eqref{rates-mod}.
Adopting the notation of \eqref{c-equations}, the inequality \eqref{entropy-bound-2} reads
\begin{align}
2 \sum_{(i,j)} m(i,j) C(i,j) \sinh(C(i,j)) \geq 2 \sum_{(i,j)} m(i,j) \big\vert \sinh(C(i,j)) \big\vert \artanh \Bigg( \frac{\sum_{(i,j)} m(i,j) \big\vert \sinh(C(i,j)) \big\vert}{\sum_{(i,j)} m(i,j) \cosh(C(i,j)) } \Bigg),
\end{align}
where we defined the symmetric matrix $m(i,j) = \sqrt{p(i) p(j)} k(i,j) \omega(i,j)$, whose entries are positive, and the sum runs over all unequal index pairs $(i,j)$ with $i \neq j$.
This is equivalent to
\begin{align}
\tanh \Bigg( \frac{\sum_{(i,j)} m(i,j) \big\vert C(i,j) \big\vert \big\vert \sinh(C(i,j)) \big\vert}{\sum_{(i,j)} m(i,j) \big\vert \sinh(C(i,j)) \big\vert} \Bigg) \geq \frac{\sum_{(i,j)} m(i,j) \big\vert \sinh(C(i,j)) \big\vert}{\sum_{(i,j)} m(i,j) \cosh(C(i,j)) } \label{inequality-2} ,
\end{align}
where we used that the hyperbolic sine is an odd function so that $x \sinh(x) = |x| |\sinh(x)|$.
We define
\begin{align}
P(i,j) = \frac{m(i,j) \big\vert \sinh(C(i,j)) \big\vert}{\sum_{(i,j)} m(i,j) \big\vert \sinh(C(i,j)) \big\vert},
\end{align}
which is normalized probability with respect to $(i,j)$,
\begin{align}
P(i,j) \geq 0 \qquad \text{and} \qquad \sum_{i,j} P(i,j) = 1 .
\end{align}
Denoting the average with respect to this probability by $\av{\ldots}$, we can write the left-hand side of \eqref{inequality-2} as
\begin{align}
\tanh \Bigg( \frac{\sum_{(i,j)} m(i,j) \big\vert C(i,j) \big\vert \big\vert \sinh(C(i,j)) \big\vert}{\sum_{(i,j)} m(i,j) \big\vert \sinh(C(i,j)) \big\vert} \Bigg) &= \tanh \big( \av{\vert C \vert} \big) \\
& \geq \Av{\tanh(\vert C \vert)} = \frac{\sum_{(i,j)} m(i,j) \big\vert \sinh(C(i,j)) \big\vert \big\vert \tanh(C(i,j)) \big\vert}{\sum_{(i,j)} m(i,j) \big\vert \sinh(C(i,j)) \big\vert} \n ,
\end{align}
where we used that the hyperbolic tangent is a concave function for positive arguments and employed Jensen's inequality.
Plugging this into \eqref{inequality-2}, we have to show that
\begin{align}
\sum_{(i,j)} m(i,j) \frac{\big\vert \sinh(C(i,j)) \big\vert^2}{\cosh(C(i,j))} \geq \frac{\Big(\sum_{(i,j)} m(i,j) \big\vert \sinh(C(i,j)) \big\vert \Big)^2}{\sum_{(i,j)} m(i,j) \cosh(C(i,j)) }.
\end{align}
However, this is nothing but the Cauchy-Schwarz inequality, for we can write,
\begin{align}
\Bigg(\sum_{(i,j)} m(i,j) \big\vert \sinh(C(i,j)) \big\vert \Bigg)^2 &= \Bigg(\sum_{(i,j)} \frac{ \sqrt{m(i,j)} \big\vert \sinh(C(i,j)) \big\vert}{\sqrt{\cosh(C(i,j))}} \sqrt{m(i,j) \cosh(C(i,j))} \Bigg)^2 \\
& \leq \sum_{(i,j)} \frac{ m(i,j) \big\vert \sinh(C(i,j)) \big\vert^2}{\cosh(C(i,j))} \sum_{(i,j)} m(i,j) \cosh(C(i,j)) \n .
\end{align}
Thus, we have proven \eqref{entropy-bound-2} by means of elementary inequalities.

\section{Relation to the minimum entropy production rate of Ref.~\cite{Rem21}} \label{app-minent}
In Ref.~\cite{Rem21}, the entropy production rate was minimized under the constraint that the symmetric part $\omega(i,j)$ of the rates \eqref{rates} remains fixed, i.~e.~the entropy production rate was minimized only with respect to the antisymmetric part $A(i,j)$.
While, intuitively, this is a stronger constraint than fixing the activity, the relation between the resulting minimum entropy production rates has to be determined explicitly.
A central result of Ref.~\cite{Rem21} is that, when fixing the symmetric part of the rates, the optimal forces $A(i,j)$ that minimize the entropy production rate do generally not satisfy the detailed balance condition \eqref{db-rates}.
Let us consider an alternative minimization problem.
Instead of the entropy production rate, we minimize the functional
\begin{align}
g = \sigma - 2 \chi
\end{align}
under the constraint that the time evolution of the occupation probabilities and the symmetric part $\omega(i,j)$ are kept fixed.
This is similar to the problem considered in Ref.~\cite{Rem21}, the difference being that we subtract twice the activity from the entropy production rate.
We recall the notation introduced in \eqref{c-equations}
\begin{subequations}
\begin{align}
d_t p(i) &= 2 \sum_j \sqrt{p(i) p(j)} k(i,j) \omega(i,j) \sinh(C(i,j)), \label{c-equations-app-1} \\
\sigma &= 2 \sum_{i,j} \sqrt{p(i) p(j)} k(i,j) \omega(i,j) C(i,j) \sinh(C(i,j)), \label{c-equations-app-2}  \\
\chi &= \sum_{i,j \neq i} \sqrt{p(i) p(j)} k(i,j) \omega(i,j) \cosh(C(i,j)) \label{c-equations-app-3} .
\end{align}\label{c-equations-app}%
\end{subequations}
Minimizing $g$ with respect to $A(i,j)$ is equivalent to minimizing with respect to $C(i,j)$ since the two are related via the linear transformation \eqref{a-c-relation}.
We obtain the condition for a stationary point
\begin{align}
k(i,j) \omega(i,j) C(i,j) \cosh(C(i,j)) = k(i,j) \omega(i,j) \big(\lambda(j) - \lambda(i) \big) \cosh(C(i,j)),
\end{align}
which implies that, whenever $k(i,j) \omega(i,j) > 0$, we should have
\begin{align}
C(i,j) = \lambda(j) - \lambda(i) .
\end{align}
This means that the transition rates that minimize $g$ have to satisfy the detailed balance condition.
Mathematically, subtracting the term $2\chi$ precisely cancels the term that leads to a breaking of detailed balance in Ref.~\cite{Rem21}.
Now suppose that we have found the rates that minimize $g$.
We denote the corresponding entropy production $\sigma_{g\vert \omega}$ and activity $\chi_{g\vert \omega}$ with a subscript $g\vert \omega$ indicating that they were obtained by minimizing $g$ while fixing $\omega$.
Since the rates satisfy detailed balance, the result of Ref.~\cite{Rem21} implies that there exists another set of rates, possessing the same symmetric part and the same time evolution, which do not satisfy detailed balance and lead to a smaller entropy production rate $\sigma_{\sigma \vert \omega} < \sigma_{g \vert \omega}$.
We denote the corresponding activity by $\chi_{\sigma \vert \omega}$.
However, since we minimized $g$, we have the inequality
\begin{align}
g_{g \vert \omega} = \sigma_{g \vert \omega} - 2 \chi_{g \vert \omega} \leq \sigma_{\sigma \vert \omega} - 2 \chi_{\sigma \vert \omega} = g_{\sigma \vert \omega} .
\end{align}
Since $\sigma_{\sigma \vert \omega} < \sigma_{g \vert \omega}$, this implies
\begin{align}
\chi_{\sigma \vert \omega} < \chi_{g \vert \omega} .
\end{align}
So the minimum entropy production dynamics of Ref.~\cite{Rem21} necessarily also has a lower activity than the dynamics minimizing $g$.
Now we fix the activity $\chi_{\sigma \vert \omega}$ and again minimize the entropy production rate according to Section \ref{sec-minent}.
We thus obtain a yet lower minimum entropy production
\begin{align}
\sigma^*_{\sigma \vert \omega} = 2 \chi^* \artanh \bigg(\frac{\chi^*}{\chi_{\sigma \vert \omega}} \bigg) .
\end{align}
Note that since $\chi^*$ only depends on the time evolution of the occupation probabilities and the connectivity of the state network, it is the absolute minimum value of the activity and independent of whether we start from the original dynamics or the minimum entropy production dynamics at fixed $\omega$.
Since we have $\chi_{\sigma \vert \omega} < \chi_{g \vert \omega}$, we have the inequalities
\begin{align}
\sigma_{\sigma \vert \omega} \geq \sigma^*_{\sigma \vert \omega} \geq \sigma^*_{g \vert \omega} .
\end{align}
Thus, starting from the detailed balanced dynamics minimizing $g$, the procedure of Section \ref{sec-minent} always yields an entropy production rate that is smaller than the one of the non-detailed balanced dynamics minimizing the entropy production rate for fixed $\omega$.
However, when compared to the original dynamics, we have
\begin{align}
\sigma_{\sigma \vert \omega} \geq \sigma^*_{\sigma \vert \omega} = 2 \chi^* \artanh \bigg(\frac{\chi^*}{\chi_{\sigma \vert \omega}} \bigg) \lessgtr 2 \chi^* \artanh \bigg(\frac{\chi^*}{\chi} \bigg) = \sigma^* \label{minent-inequality},
\end{align}
since the minimum entropy production dynamics of Ref.~\cite{Rem21} does not always lead to a smaller activity compared to the original dynamics.
As a consequence, either one of the minimum entropy production values $\sigma^*$ and $\sigma_{\sigma \vert \omega}$ may be the smaller one.
However, we expect the relation $\sigma_{\sigma \vert \omega} \geq \sigma^*$ to hold in most cases.
The reason for this becomes apparent when looking at \eqref{c-equations-app}:
Both the entropy production rate and the activity are convex functionals of the parameters $C(i,j)$.
Thus, minimizing the entropy production rate corresponds to choosing the parameters $C(i,j)$ in \eqref{c-equations-app-2} as small as possible, while still satisfying \eqref{c-equations-app-1}.
Except for special cases, this choice should also reduce the value of the activity, \eqref{c-equations-app-3}.
Thus, we may expect $\chi_{\sigma \vert \omega} \leq \chi$ and thus $\sigma_{\sigma \vert \omega} \geq \sigma^*$ via \eqref{minent-inequality}.

\section{Minimum entropy production dynamics for finite time} \label{app-finite}
In \eqref{minent-finite}, we argued that the minimum entropy production for the process connecting an initial probability $\bm{p}_\text{i}$ to a final probability $\bm{p}_\text{f}$ is realized by a process with a constant rate of change of the probability.
Here, we want to show this fact explicitly.
We can divide the problem of minimizing the entropy production into two steps.
First, for a given time evolution of the probability $\bm{p}_t$ and activity $\chi_t$, we minimize the entropy production rate at each instant of time.
This leads to the value of $\sigma_t^*$ given by \eqref{minent}.
Second, we minimize the resulting entropy production
\begin{align}
\Delta S^\text{irr} = \int_0^\tau dt \ \sigma_t^* = 2 \int_0^\tau dt \ \chi_t^* \artanh\bigg(\frac{\chi_t^*}{\chi_t} \bigg) \label{minent-local}.
\end{align}
under the constraints that $\bm{p}_0 = \bm{p}_\text{i}$ and $\bm{p}_\tau = \bm{p}_\text{f}$ and that the total number of transitions is given by
\begin{align}
\av{M} = \int_0^\tau dt \ \chi_t .
\end{align}
We note that, as discussed in Section \ref{sec-wasserstein} the minimum activity $\chi_t^*$ depends only on the time-derivative of the probability vector $d_t \bm{p}_t$.
Next, we use \eqref{minent-mod}, which allows us to construct a dynamics with $\chi_t = \mathcal{A}_t \chi_t^*$, where the choice of $\mathcal{A}_t \geq 1$ is arbitrary.
This means that we can treat $\chi_t$ and $\chi_t^*$ as independent.
Then, we have from \eqref{minent-local},
\begin{align}
\Delta S^\text{irr} = \int_0^\tau dt \ \sigma_t^* = 2 \int_0^\tau dt \ \chi_t^* \artanh\bigg(\frac{1}{\mathcal{A}_t} \bigg)
\end{align}
with the constraint on the activity now being
\begin{align}
\av{M} = \int_0^\tau dt \ \mathcal{A}_t \chi^*_t .
\end{align}
First, we fix $\chi_t^*$, and, from the Euler-Lagrange equation for the minimization with respect to $\mathcal{A}_t$ we find
\begin{align}
2 \artanh\bigg(\frac{1}{\mathcal{A}_t} \bigg) + \lambda \mathcal{A}_t = 0 ,
\end{align}
where $\lambda$ is a Lagrange multiplier.
This condition implies that $\mathcal{A}_t \equiv \mathcal{A}$ has to be independent of time.
Plugging this into the above equations, we obtain
\begin{align}
\Delta S^\text{irr} = 2 \artanh\bigg(\frac{1}{\mathcal{A}} \bigg) \int_0^\tau dt \ \chi_t^*  \qquad \text{and} \qquad \av{M} = \mathcal{A} \int_0^\tau dt \ \chi^*_t ,
\end{align}
and, solving the second equation for $\mathcal{A}$,
\begin{align}
\Delta S^\text{irr} = 2 \int_0^\tau dt \ \chi_t^* \ \artanh\bigg(\frac{\int_0^\tau dt \ \chi_t^*}{\av{M}} \bigg) .
\end{align}
This is a monotonously increasing function of $\int_0^\tau dt \ \chi_t^*$, so minimizing it requires minimizing $\int_0^\tau dt \ \chi_t^* = \int_0^\tau \ \mathcal{W}(\bm{p}_{t+dt},\bm{p}_t)$ under the constraint of the initial and final probability.
However, we already saw in Section \ref{sec-finite-time} that a constant rate of change of the probability defines the geodesic in Wasserstein space, so the former indeed minimizes the entropy production and we obtain \eqref{minent-finite}.


%

\end{document}